\documentclass[10pt]{article}

\usepackage{amsmath}
\usepackage{array}
\usepackage{appendix}
\usepackage{graphicx}
\usepackage{amsfonts}
\usepackage{amssymb}
\usepackage{mathrsfs}
\usepackage{yfonts}
\usepackage{euscript}
\usepackage{upgreek}
\usepackage{slantsc}
\usepackage{calligra}
\usepackage[T1]{fontenc}
\usepackage{epsf}
\usepackage{latexsym}

\usepackage{tipa}

\def\be{\begin{equation}}
\def\ee{\end{equation}}
\def\beq{\begin{equation}}
\def\eeq{\end{equation}}
\def\bea{\begin{eqnarray}}
\def\eea{\end{eqnarray}}

\def\foo{\footnote}

\def\hat{\widehat}

\def\!{\hspace{-0.3em}}

\def\mD{\mbox{D}}

\def\mF{\mbox{F}}

\def\mH{\mbox{H}}

\def\mP{\mbox{P}}

\def\mS{\mbox{S}}

\def\ma{\mbox{a}}

\def\md{\mbox{d}}

\def\mg{\mbox{g}}

\def\mk{\mbox{k}}
\def\ml{\mbox{l}}

\def\mt{\mbox{t}}

\def\Bigalpha{\mbox{\Large $\alpha$}}               
\def\Bigbeta{\mbox{\Large $\beta$}}                
\def\Biggamma{\mbox{\Large $\gamma$}}               

\def\Bigeta{\mbox{\Large $\eta$}}

\def\brho{\mbox{\boldmath$\rho$}}          
\def\bpi{\mbox{\boldmath$\pi$}}            %
\def\bdelta{\mbox{\boldmath$\delta$}}

\def\sbSigma{\mbox{\scriptsize\boldmath$\Sigma$}} 

\def\btheta{\mbox{\boldmath$\theta$}}

\def\fa{\mbox{\sffamily a}}

\def\fD{\mbox{\sffamily D}}

\def\fG{\mbox{\sffamily G}}

\def\fP{\mbox{\sffamily P}}
\def\fQ{\mbox{\sffamily Q}}

\def\fS{\mbox{\sffamily S}}

\def\ip{i^{\prime}}

\def\bB{\mbox{\bf B}}

\def\bM{\mbox{\bf M}}
\def\bl{\mbox{\bf l}}
\def\bQ{\mbox{\bf Q}}
\def\bS{\mbox{\bf S}}

\def\bM{\mbox{{\bf M}}}
\def\bM{\mbox{{\bf M}}}

\def\scE{\mbox{\scriptsize ${\cal E}$}}          

\def\scc{\mbox{\scriptsize c}}
\def\sd{\mbox{\scriptsize d}}
\def\se{\mbox{\scriptsize e}}

\def\si{\mbox{\scriptsize i}}
\def\sj{\mbox{\scriptsize j}} 
\def\sk{\mbox{\scriptsize k}}
\def\sll{\mbox{\scriptsize l}}  
\def\sm{\mbox{\scriptsize m}}
\def\sn{\mbox{\scriptsize n}} 
\def\so{\mbox{\scriptsize o}} 
\def\sp{\mbox{\scriptsize p}}

\def\sr{\mbox{\scriptsize r}}
\def\sss{\mbox{\scriptsize s}}  
\def\st{\mbox{\scriptsize t}}

\def\sB{\mbox{\scriptsize B}}
\def\sC{\mbox{\scriptsize C}}

\def\sG{\mbox{\scriptsize G}}

\def\sJ{\mbox{\scriptsize J}}
\def\sK{\mbox{\scriptsize K}}
 
\def\sM{\mbox{\scriptsize M}} 
 
\def\sO{\mbox{\scriptsize O}}
\def\sP{\mbox{\scriptsize P}} 
 
\def\sR{\mbox{\scriptsize R}}
\def\sS{\mbox{\scriptsize S}}

\def\sU{\mbox{\scriptsize U}}

\def\sW{\mbox{\scriptsize W}}

\def\sfa{\mbox{\sffamily{\scriptsize a}}}
\def\sfb{\mbox{\sffamily{\scriptsize b}}}
\def\sfc{\mbox{\sffamily{\scriptsize c}}}
\def\sfd{\mbox{\sffamily{\scriptsize d}}}
\def\sfe{\mbox{\sffamily{\scriptsize e}}}
\def\sff{\mbox{\sffamily{\scriptsize f}}}

\def\sfp{\mbox{\sffamily{\scriptsize p}}}
\def\sfq{\mbox{\sffamily{\scriptsize q}}}

\def\sfs{\mbox{\sffamily{\scriptsize s}}}

\def\sfA{\mbox{\sffamily{\scriptsize A}}}
\def\sfB{\mbox{\sffamily{\scriptsize B}}}

\def\sfG{\mbox{\sffamily{\scriptsize G}}}

\def\sfP{\mbox{\sffamily{\scriptsize P}}}

\def\sfS{\mbox{\sffamily{\scriptsize S}}}

\def\sfZ{\mbox{\sffamily{\scriptsize Z}}}

\def\sbM{\mbox{{\bf \scriptsize M}}}

\def\sbU{\mbox{{\bf \scriptsize U}}}

\def\ta{\mbox{\tiny a}}

\def\te{\mbox{\tiny e}}

\def\th{\mbox{\tiny h}}
\def\ti{\mbox{\tiny i}}

\def\tl{\mbox{\tiny l}}

\def\tm{\mbox{\tiny m}}
\def\tn{\mbox{\tiny n}}

\def\tp{\mbox{\tiny p}}

\def\tr{\mbox{\tiny r}}
\def\ts{\mbox{\tiny s}}

\def\tU{\mbox{\tiny U}}

\def\tfA{\mbox{\sffamily{\tiny A}}}

\def\tfZ{\mbox{\sffamily{\tiny Z}}}

\def\ttM{\mbox{\tt{M}}}

\def\lt{\mbox{\Large $t$}}

\def\K{Kucha\v{r} }

\def\pa{\partial}
\def\d{\textrm{d}}

\def\partional{\bdelta\hspace{-0.08in}\pa}    
\def\Last{\mbox{\Large$\ast$}}                
\def\5Star{\mbox{\Large$\star$}}              
\def\Rec{\mbox{\textcircled{$\star$}}}        

\def\Diamond{\mbox{\Large$\diamond$}}          
\def\Heart{\mbox{$\heartsuit$}}            
\def\Spade{\mbox{$\spadesuit$}}           
\def\Club{\mbox{$\clubsuit$}}             

\def\SStar{\mbox{\normalsize$\ast$}}

\def\cr{\mbox{\scriptsize{\bf $\mbox{ } \times \mbox{ }$}}}

\def\sumi3{\sum\mbox{}_{\mbox{}_{\mbox{\scriptsize $i$=1}}}^3}

\def\sumj3{\sum\mbox{}_{\mbox{}_{\mbox{\scriptsize $j$=1}}}^3}
\def\sumk3{\sum\mbox{}_{\mbox{}_{\mbox{\scriptsize $k$=1}}}^3}

\begin{document}

\begin{titlepage}

\normalfont

\vspace{.7in}
\begin{center}
\Large{\bf MACHIAN CLASSICAL AND SEMICLASSICAL EMERGENT TIME}

\vspace{.1in}

\normalsize

\vspace{.4in}

{\large \bf Edward Anderson}

\vspace{.2in}

\large {\em DAMTP, Centre for Mathematical Sciences, Wilberforce Road, Cambridge CB3 OWA.  } \normalsize

\end{center}

\vspace{1in}

\begin{abstract}

Classical and semiclassical schemes are presented that are timeless at the primary level and recover time from 
Mach's `time is to be abstracted from change' principle at the emergent secondary level.
The semiclassical scheme is a Machian variant of the Semiclassical Approach to the Problem of Time in Quantum Gravity. 
The classical scheme is Barbour's, cast here explicitly as the classical precursor of the Semiclassical Approach. 
Thus the two schemes have been married up, as equally-Machian and necessarily distinct, since the latter's timestandard is abstracted in part from quantum change.  
I provide perturbative schemes for these in which the timefunction is to be determined rather than assumed.  
This paper is useful modelling as regards the Halliwell--Hawking arena for the quantum origin of the inhomogeneous cosmological fluctuations.

\end{abstract}

\vspace{1in}

          		     		   
\vspace{3in}

\noindent $^*$ ea212@cam.ac.uk   

\end{titlepage}

\section{Introduction}\label{Intro}

\subsection{Resolution of three facets of the classical Problem of Time}

\noindent This account of Physics starts by considering {\it configuration space} $\fQ$, i.e. the space of all possible configurations $Q^{\sfA}$ that a physical system can take. 
In ordinary mechanics, the configurations are particle positions \cite{Lanczos}. 
In field theories they are the values taken by the field on a fixed spatial slice. 
In GR they are the values taken by the 3-metrics on a spatial slice with fixed spatial topology, $\Sigma$ \cite{DeWitt67}.  
One then builds composite objects from the configurations, one's first goal being to write down an action for one's theory.  
(Other composite objects include notions of distance, of information and of correlation \cite{FileR}.)

{\it Temporal Relationalism} \cite{B94I, SemiclI, ARel, FileR} is then the classical precursor of the well-known Frozen Formalism Problem facet of the Problem of Time.
It concerns the Leibnizian `no time for the universe as a whole' idea \cite{B94I, EOT, FileR}.  
This is mathematically implemented by use of geometrical actions that happen to be parametrization-irrelevant,\footnote{  
$\bM$ is the metric on configuration space. 
For Mechanics, $W := E - V$, where $V$ is the potential energy and $E$ is the total energy.
The $P_{\tfA}$ are the momenta conjugateto the $Q^{\tfA}$.
For GR (minisuperspace models for now), $W := 2\Lambda$ -- Ric and $\bM$ is (a truncation of) the inverse undensitized GR supermetric, 
$M^{\mu\nu\rho\sigma} = h^{\mu\rho}h^{\nu\sigma} - h^{\mu\nu}h^{\rho\sigma}$.}  
\beq
S_{\sr\se\sll} = \sqrt{2}\int\d \widetilde{s} = \sqrt{2}\int\d s\sqrt{W(\bQ)} \mbox{ } \mbox{ for } \d s := ||\d\bQ||_{\sbM} = \sqrt{M_{\sfA\sfB}\d Q^{\sfA}\d Q^{\sfB}} \mbox{ } . 
\label{Srel}
\eeq
The first form involves the physical line element $\d\widetilde{s}$. 
On the other hand, the second expression contains the conformally-related configuration space geometrical line element $\d s$.  
For mechanics, (\ref{Srel}) is Jacobi's formulation \cite{Lanczos}, and for minisuperspace it is Misner's formulation \cite{Magic}.
Parametrization-irrelevant actions must lead, by Dirac's argument \cite{Dirac}, to primary constraints.  
These include the well-known Hamiltonian constraint of GR and the energy constraint of relational particle mechanics (RPM) models. 
Both of these constraints are purely quadratic in the corresponding momenta due to the square-root form of $\d s$.

The above timelessness is then to be resolved by Mach's `time is to be abstracted from change'. 
Three alternatives for this involve  `any change' (Rovelli \cite{Rfqxi}), `all change' (Barbour \cite{Bfqxi}), or my sufficient totality of locally significant change (STLRC) \cite{ARel2}.  
In the last case, a generalized local ephemeris time (GLET) \cite{ARel2} emerges.
%
%
To fulfil the true content of the STLRC implementation of Mach's Time Principle, all change is given opportunity to contribute to the timestandard. 
However only changes that do so in practise to within the desired accuracy are actually kept. 
Moreover, this approximation requires a curious indirect procedure.  
I.e. one can not simply compare the sizes of the various contributions to the energy equation. 
One must rather \cite{ARel2} assess terms at the level of the resulting force terms that arise upon variation.

The emergent Jacobi--Barbour--Bertotti (JBB) time (in a form suitable for Mechanics or minisuperspace) is given by
\be
{\d}/{\d t^{\se\sm(\sJ\sB\sB)}} = \sqrt{2W(\bQ)}{\d}/{\d s} \mbox{ or, integrating, } \mbox{ } 
\lt^{\se\sm(\sJ\sB\sB)} = \int\d s\left/\sqrt{2 W(\bQ)}\right. \mbox{ } .
\label{plain-tem}
\eeq
\noindent Here, $\mbox{\Large $t$}^{\se\sm(\sJ\sB\sB)} := t^{\se\sm(\sJ\sB\sB)} - t^{\se\sm(\sJ\sB\sB)}_{(0)}$, thus incorporating a `choice of `calendar year zero'.  
A constant scaling `constant tick length' can also be included \cite{FileR}.  
Using $\lt^{\se\sm(\sJ\sB\sB)}$ simplifies both the momentum--change relations   
$
P_{\sfA} = \d \, \d\widetilde{s}/\d\d Q^{\sfA} = \sqrt{2W}M_{\sfA\sfB}\d Q^{\sfB}/\d s
$
and the Jacobi counterpart of the Euler--Lagrange equations of motion,     
$
\d P_{\sfA} = \d \, \d \widetilde{s}/\d Q^{\sfA} = \d \sqrt{2W} \d s/\d Q^{\sfA} \mbox{ } .  
$  
Moreover, $\lt^{\se\sm(\sJ\sB\sB)}$ leads to a relational recovery of what is, in various suitable contexts, Newtonian time, proper time and cosmic time. 
Finally, $\lt^{\se\sm(\sJ\sB\sB)}$ is, on first sight, built from an `all change' expression, but, upon practical consideration \cite{ARel2, FileR}, it is a STLRC.
Thus this timestandard itself is a local generalization of the astronomers' ephemeris time \cite{Clemence}.   
Explicit forms for this have been worked out for 1- and 2-$d$ RPM's \cite{FileR, QuadI} and for minisuperspace \cite{AMSS1}.


\noindent This resolution of the Frozen Formalism Problem facet of the Problem of Time then meets two complications. 


\noindent 1) The emergent JBB time fails to unfreeze the quantum wave equation.  

\noindent The riposte to 1) is to consider a Machian semiclassical approach which gives rise to a semiclassical Machian timestandard $\lt^{\se\sm(\sW\sK\sB)}$.  
Moreover this timestandard is indeed {\sl expected} to be be different from it on Machian grounds.
This is because there  there are now quantum, rather than classical, light ($l$)-degrees of freedom to be given the opportunity to contribute.   
\cite{FileR} and the present article are the first to comment on the extent to which the semiclassical approach is 
a) Machian and b) has a well-known (and also Machian) classical precursor.  

\noindent 2) A second facet interferes.  In the case of classical GR, this most usually termed the Thin Sandwich Problem \cite{Kuchar92, I93}. 
Moreover, it generalizes to a wider range of theories as Barbour's Best Matching Problem. 
It furthermore generalizes as regards at which level it is tackled.  
The Thin Sandwich is specifically at the `Lagrangian' level, or, in the fully relational formulation, at the Jacobi level: in terms of $Q^{\sfA}$, d$Q^{\sfA}$ variables. 
On the other hand, {\it Configurational Relationalism} can be at other levels, such as the classical Hamiltonian level or at some quantum level.   
The interference of this second facet is clear from the action now containing auxiliary $\fG$-variables for $\fG$ a group of physically irrelevant transformations.  
Moreover, one now takes one's emergent time object to be a $\fG$-extremization of one's action,\footnote{Examples of this are the RPM action in the next SSec and the BFO-A 
(Barbour--Foster--\'{o} Murchadha \cite{RWR}) 
action of Baierlein--Sharp--Wheeler \cite{BSW} type for GR: here $\fG$ = Diff($\Sigma$) and the $\mg_{\tfZ}$ are presented in the form $F^{\mu}$ (frame variables).}
\beq
S_{\sr\se\sll} = \sqrt{2}\int_{\Sigma}\d\Sigma\int||\d_{g}\bQ||_{\sbM}\sqrt{W(\bQ)}  
\label{tJBB}
\eeq
[$\int_{\Sigma}\d\Sigma$ is taken to be 1 for RPM and minisuperspace.]  
Then
\beq
\lt^{\se\sm(\sJ\sB\sB)} = \stackrel{\mbox{\scriptsize extremum }}
                                    {\mbox{\scriptsize $\d g \in $ $\sfG$ of $S_{\tr\te\tl}$}}
\left( \int||\d_{g}\bQ||_{\sbM}/\sqrt{2 W(\bQ)}\right) \mbox{ } . 
\label{G-tem}
\eeq 
N.B. (\ref{tJBB})'s local character: GR time is a function of local position. 
Moreover, for 1- and 2-$d$ RPM's  \cite{FileR} {\sl and for the below Halliwell--Hawking arena that they model} \cite{forth}, this expression is explicitly evaluable 
via Best Matching/Thin Sandwich being resolved.

The above also ensures a set of classical \K beables, thus also resolving a third Problem of Time facet \cite{Kuchar92, I93, APoT, APoT2}: the Problem of Beables.
Observables, or beables  -- following John Bell \cite{Bell}: a more cosmologically-appropriate notion than observables -- are hard to come by in classical and quantum GR. 
Resolved Best Matching readily implies possession of a full set of classical \K beables, i.e. quantities that Poisson-brackets-commute with the classical linear constraints.    
For more on the Problem of Time, see \cite{Kuchar92, I93, APoT2}.

\subsection{Relational Particle Mechanics (RPM)}

This is an example of nontrivial Configurational Relationalism \cite{BB82}, and is the main concrete example used in this paper. 
The action for scaled RPM is\footnote{$\rho^{i\mu}$ are mass-weighted Jacobi 
inter-particle cluster relative coordinates \cite{Marchal} with conjugate momenta $\pi_{i\mu}$.
These are the most convenient relative coordinates due to their diagonalizing the kinetic term.  
$B^{\mu}$ is a rotational auxiliary variable.
The lower-case Latin letters are relative particle (cluster) labels running from 1 to $n$ = $N$ -- 1 for $N$ the number of particles.
$W = E_{\tU\tn\ti} - V$, for potential $V$ and fixed total energy of the model universe $E_{\tU\tn\ti}$.
The $\btheta$ are general $n$-sphere coordinates and $\Theta$, $\Phi$ are spherical coordinates on triangleland (the 3-particle RPM in dimension 2 or higher). 
I.e. a function of the ratio between the base and the median and the angle between the base and the median respectively.
As explained in \cite{QuadI}, much of the notation and concepts for RPM's come from theoretical Molecular Physics' kinematics and from Kendall-type Shape Theory \cite{Kendall}.}
\beq
S_{\sR\sP\sM} = \sqrt{2}\int\sqrt{W}\d_B s \mbox{ } , \mbox{ } \mbox{ } \d_B s := ||\d{\brho} - \d  \bB \cr \brho|| \mbox{ } .  
\label{Uuno}
\eeq
\noindent The quadratic energy constraint  
\beq
{\cal E} := ||\bpi||^2/2 + V(\brho) = E_{\sU\sn\si} \mbox{ }     
\label{EnCon}
\eeq
then follows as a primary constraint, i.e. purely due to the form of the action with no variation performed \cite{Dirac}. 
${\cal E}$ and GR's Hamiltonian constraint, ${\cal H}$, are denoted collectively by Quad, which emphasizes their quadraticity in the momenta.  
Furthermore, variation with respect to the auxiliary $G$ variables produces a zero total angular momentum constraint that is linear in the momenta,   
\beq
\underline{\cal L} := \sum\mbox{}_{\mbox{}_{\mbox{\scriptsize $i = 1$}}}^n \brho^i \cr \bpi_i = 0 \mbox{ } .
\label{ZAM}
\eeq 
The specific examples of RPM's used in this paper are all scaled: they are $N$-stop metroland ($N$ particles on a line), in particular 3-stop metroland, and triangleland.\footnote{This  
nomenclature is necessary since these are {\sl not} the same as $N$-body problems. 
The latter carry implications of being a small subsystem within a larger universe whereas ours are whole-universe models.  
This leads to mathematical differences between the two at the quantum level \cite{FileR}.}
%
Reduction can be performed for these (and in fact for all $N$-a-gonlands and for all the pure-shape -- i.e. shape alone and not scale -- versions of all of these also). 
Equivalently by \cite{FORD, Cones, FileR}, one can set up a mechanics on the configuration space geometry. 
I refer to the common outcome of these two procedures as the {\it r-formulation}. 
In the case of $N$-stop metroland, the action is 
\beq
S_{N-\sss\st\so\sp} = \sqrt{2}\int\sqrt{W(\rho, \btheta)}\d s \mbox{ } , \mbox{ } \mbox{ } \d s := \sqrt{\d\rho^2 + \rho^2\d s_{\mathbb{S}^{N - 2}}^2(\btheta)} \mbox{ } ,
\eeq
corresponding to the configuration space geometry being $\mathbb{R}^{N - 1}$.
On the other hand, for triangleland, 
\beq
S_{\triangle} = \sqrt{2}\int\sqrt{W(\Theta,\Phi)}\d s \mbox{ } , \mbox{ } \mbox{ } \d s := \sqrt{   \{   \d I^2 + I^2\d s_{\mathbb{S}^{2}}^2(\Theta, \Phi)   \}/4I   } \mbox{ } ,
\eeq
corresponding to the configuration space geometry being $\mathbb{R}^3$ with a non-flat (but conformally flat) metric.
Here, $I$ is the total moment of inertia of the model universe.  
The advantages of considering triangleland are that it incorporates nontrivial configurational relationalism. 
%

\noindent RPM's generalize previously-studied absolute particle models of the Semiclassical Approach 
\cite{Banks, BriggsRost, Datta97, Pad, PadSingh} by inclusion of auxiliary terms and subsequently of linear constraints.    
See \cite{AMSS1} for a minisuperspace model arena treatment of the present paper's approach.

\subsection{Motivation: qualitative study of robustness of Halliwell--Hawking model}

\noindent I provide below an overview of the standard Semiclassical Approach.  
Suppose that \cite{DeWitt67, LR79, Banks, HallHaw, Kuchar92, I93, Kieferbook} there are slow, heavy `$h$' variables that provide an approximate timestandard with respect to which 
                                                                              the other fast, light `$l$' degrees of freedom evolve.    
The Semiclassical Approach is not only an emergent time strategy toward resolving the Problem of Time.  
It is also used along the lines of e.g. Halliwell and Hawking \cite{HallHaw}) in acquiring more solid foundations for other aspects of Quantum Cosmology.  
This Halliwell--Hawking model is for the quantum-cosmological origin of cosmological fluctuations observed today (microwave background hot-spots and galaxies).
These are treated as small inhomogeneous perturbations ($l$) about the spatially homogeneous and isotropic $\mathbb{S}^3$ universe ($h$).
The present paper studies RPM's as a simpler model arena for understanding the Halliwell--Hawking model.  
In particular, I give a qualitative study of a more general perturbation scheme that is to be a robustness test for the Halliwell--Hawking model 
(where many terms were droppod from the equations with little comment).
RPM's then already succeed in illustrating the qualitative differences upon keeping these various terms, 
whilst having equations that are around 5 times shorter and more tractable than the corresponding generalization of the Halliwell--Hawking model.
[This factor of 5 comes from the Halliwell--Hawking model having scalar, vector and tensor gravitational modes and inhomogeneous scalar field modes, 
which form a total of five singlet or even--odd doublet modes.]

\subsection{The standard Semiclassical Approach}

I concentrate in this article on the case of particular cosmological significance: scale = $h$, shape = $l$ splits. 
For GR, $h$ is the scalefactor (and homogeneous matter modes) and $l$ are inhomogeneities (treated perturbatively in the Halliwell--Hawking scheme \cite{HallHaw}).  
Whereas for RPM's $h$ is (the square root of) the moment of inertia for the whole universe and $l$ are pure-shape degrees of freedom.  
The Semiclassical Approach then involves making

\noindent i) the Born--Oppenheimer and WKB ans\"{a}tze are, respectively, 
\beq
\Psi(h, l) = \psi(h)|\chi(h, l)\rangle \mbox{ } ,
\label{BO}
\eeq 
\beq
\psi(h) = \mbox{exp}(i\,S(h)/\hbar) \mbox{ } .
\label{WKB}
\eeq 
(In each case there are a number of associated approximations covered in Secs 2, 3 and Appendix A.)    

\noindent ii) One forms the $h$-equation, 
$
\langle\chi| \widehat{\mbox{Quad}} \, \Psi = 0.  
$  
Then, under a number of simplifications (including, for later reference, neglect of the light subsystem's kinetic term, $T_l$) this yields a Hamilton--Jacobi equation\foo{For simplicity, I 
present this in the case of 1 $h$ degree of freedom and with no linear constraints.  
More generally it involves contracted into inverse kinetic metric $N$ and an accompanying linear constraint, see Sec \ref{Semicl} for these in detail.
$N^{hh}$ is the $hh$-component of the inverse of the configuration space metric.} 
$N^{hh}\{\pa S/\pa h\}^2 = 2\{E - V(h)\}$ where $V(h)$ is the $h$-part of the potential. 
One way of solving this is for an approximate emergent semiclassical time, $t^{\se\sm(\sW\sK\sB)}(h)$. 

\noindent iii) Next, one forms the $l$-equation, 
$
\{1 - |\chi\rangle\langle\chi|\}\widehat{\mbox{Quad}}\,\Psi = 0. 
$
This fluctuation equation can be recast (modulo further approximations) into an emergent-WKB-TDSE (time-dependent Schr\"{o}dinger equation) for the l-degrees of freedom. 
The mechanics/RPM form of this is 
\beq
i\hbar\pa|\chi\rangle/\pa t^{\se\sm(\sW\sK\sB)}  = \widehat{\cal E}_{l}   |\chi\rangle \mbox{ } .
\label{TDSE2}
\eeq
The emergent-time-dependent left-hand side arises from the cross-term $\pa_{h}|\chi\rangle\pa_{h}\psi$. 
$\widehat{\cal E}_{l}$ is the remaining piece of the quantum energy constraint $\widehat{\cal E}$, acting as a Hamiltonian for the $l$-subsystem.  
\noindent We shall see there are similar forms to (\ref{plain-tem}, \ref{G-tem}) for $t^{\se\sm(\sW\sK\sB)}$.

\subsection{Outline of the rest of this paper}

\noindent Sec \ref{+temJBB}'s Machian classical $h$--$l$ split (Level 1 of the current program) is furtherly motivated as a simplification of semiclassical scheme 
associated with well-known physics.   
It is already Machian, and is Level 1 of the current program.
\noindent Sec \ref{Semicl} gives more detail of the semiclassical approach, including of how it too can be cast in Machian form. This is Level 2 of the current program.  
\noindent N.B. that the working leading to a time-dependent Schr\"{o}dinger equation ceases to work in the absence of making the WKB ansatz and approximation 
\cite{Zeh88, BV89, BS, Kuchar92, I93, B93, Kiefer94, SemiclI}.
Thus for Quantum Cosmology, this is not known to be a particularly strongly supported ansatz and approximation to make.    
Propping this up requires considering one or two further Problem of Time strategies from the classical level upwards (see the Conclusion).  
$t^{\se\sm(\sW\sK\sB)}$ aligns with $t^{\se\sm(\sJ\sB\sB)}$ at least to first approximation.  
\noindent As outlined in the Conclusion (see \cite{H03, H09, H11, AHall, FileR, CapeTown12} for more details),  
this justifies the WKB ansatz leads one to a Machian Semiclassical Histories Timeless Records combined scheme: Level 3 of the current program.     
The present article covers Levels 1 and 2.

The first approximation for the emergent time coincides in the classical and semiclassical workings.
However, this is rather un-Machian in the sense that it abstracts its change just from the scale.
(Sometimes this is alongside homogeneous isotropic matter modes, or, more widely, from the usually-small subset of $h$ degrees of freedom.) 
However, in the second approximation, the $l$ degrees of freedom {\sl are} given the opportunity to contribute to the corrected emergent time.  
Moreover, they do so differently in the classical and semiclassical cases.  
In this paper, this is an RPM pure-shape change, though in subsequent papers \cite{AMSS1, forth} it is a minisuperspace anisotropy and a perturbative GR inhomogeneity.  

\noindent Finally, the number of approximations concurrently made in the Semiclassical Approach is large (`Multiple Approximations Problem' \cite{FileR}).   
There are non-adiabaticities, other (including higher) emergent time derivatives and averaged terms \cite{SemiclI, MP98, Arce}.  
Including the last of these parallels the use of Hartree--Fock self-consistent iterative schemes.
However, in the present context the system is now more complex via involving a chroniferous (`time providing') quantum-average-corrected Hamilton--Jacobi equation.  
These have the effect of obscuring tests of the validity of the WKB approximation -- the truth involves vast numbers of different possible regimes.
Thus tests of validity are likely contingent on a whole list of approximations made.  
These approximations are covered in Secs 2 and 3 as they arise.  
I note that combined classical and quantum perturbation schemes are unusual, as are perturbation schemes for fixing the timestandard rather than built on a presupposed timestandard.

\section{Machian Classical Scheme}\label{+temJBB}

\subsection{Heavy--light ($h$-$l$) splits}

\noindent Suppose a classical system has a regime exhibiting a split is between $h$ and $l$ degrees of freedom.
This is a classical parallel of the Born--Oppenheimer \cite{BO27} split of Molecular Physics \cite{AtF}. 
There, one solves for the electronic structure under the approximation that the much heavier nuclei stay fixed.
There is then also a technically similar approximation procedure from Semiclassical Quantum Cosmology \cite{Kieferbook}.\foo{More considerations enter `$h$--$l$ splits'
than just a mass ratio $m_l/m_h = \epsilon_{hl} << 1$: I assume `sharply peaked hierarchy' conditions (all the $h$'s have similar masses $>>$ all the similar masses of the $l$'s).
The $\epsilon$'s in this Article denote small quantities.  
A corresponding `gravitational mass hierarchy' sometimes invoked in motivating such approximations is $M_{\mbox{\scriptsize Planck}} >> M_{\mbox{\scriptsize inflaton}}$.  
Another involves the single scale factor dominating over each of the anisotropic and inhomogeneous modes in GR cosmology.} 
%
In doing so, this Sec interpolates between Classical Dynamics and Sec  \ref{Semicl}'s Semiclassical Approach to the Problem of Time.

\subsection{Scale--shape $h$-$l$ split of RPM's}\label{RPM-H-L-2}

This SSSec's particular $h$-$l$ split is aligned with scale--shape split of the RPM which has parallels with e.g. the scale--inhomogeneity split in GR.  

  
\noindent The action is now 
\be
S^{\sR\sP\sM}_{\sJ\sB\sB} = \sqrt{2}\int 
\sqrt{E_{\sU\sn\si} - V_{h} - V_{l} - J}
\sqrt{\d h^2 + h^2 ||\d \mbox{\boldmath$l$}||_{\mbox{\boldmath\scriptsize$M$}_{l}}\mbox{}^2   }
\mbox{ } ,  
\label{JHL2}
\ee
(with $\underline{B}$'s hung on the d$l$'s in the uneliminated case).
Here, 
$
V_{h}  := V_{h}(h\mbox{ alone} ) = V_{\sigma}(\sigma)
$
, 
$
V_{l} := V_{l}(l^{\sfa} \mbox{ alone} ) = V_{\sS}(\mS^{\sfa} \mbox{ alone} )
$
--- i.e. a function of pure shape alone (the lower-case sans-serif indices run over the shape degrees of freedom). 
Also, 
$
J = J(h, l^{\sfa} \mbox{ alone})= J(\sigma, \mS^{\sfa} \mbox{ alone})
$
--- the interaction term.  


The conjugate momenta are now (with multi-index $\Gamma = i\mu$ and a $\underline{B}$ hung on each $\Last l$ for $\Last := \pa/\pa t^{se\sm(\sJ\sB\sB)}$ 
in the uneliminated case and $\Gamma = \fa$ in the r-formulation case).

\be
P^{h}   = \Last h^{\ip\mu} 
\mbox{ } , \mbox{ }
P_{\Gamma}^{l} = h^2 M_{\Gamma\Lambda}\Last l^{\Lambda}  
\mbox{ } .  
\label{HLmom2}
\ee
The classical energy constraint is now  
\be
{\cal E} := P_{h}^2/2 + ||\mbox{\boldmath$P$}_{l}||_{\mbox{\boldmath\scriptsize$N$}_{l}}\mbox{}^2/2 h^2 + V_{h} + V_{l} + J = E_{\sU\sn\si} \mbox{ } .
\label{yet-another-E}
\ee
In the uneliminated case, this is accompanied by the zero total angular momentum constraint 
\be
{\underline{\cal L}_{l}} = \sum\mbox{}_{\mbox{}_{\mbox{\scriptsize $\sfa = 1$}}}^{nd - 1} \mbox{ }
\underline{l}^{\sfa} \cr \underline{P}^{l}_{\sfa}  \label{HLAMSha} \mbox{ } .
\ee
The evolution equations are [in the same notation as eq. (\ref{HLmom2})] 
\beq
\Last P^{h}   =  h||\Last l||_{\mbox{\boldmath\scriptsize$M$}_l}^2 -\pa\{V_{h} + J\}/\pa h 
\mbox{ } , \mbox{ }
\Last P_{\Gamma}^{l} =  h^2 M_{\Lambda\Sigma,\Gamma} \Last l^{\Lambda}\Last l^{\Sigma} - \pa\{V_{l} + J\}/\pa l^{\Gamma}  
\mbox{ } .
\eeq
\noindent We can treat (\ref{yet-another-E}) in $Q^{\sfA}$, d$Q^{\sfA}$ variables as an equation for $t^{\se\sm(\sJ\sB\sB)}_{(0)}$ itself.
\noindent In this classical setting, it is coupled to the $l$-equations of motion.
Furthermore, as explained in Sec \ref{iudex}, we need the $h$-equation of motion to judge which terms to keep.
If there is more than one $h$ degree of freedom, there is separate physical content in these from that of the energy equation.    
The system is in general composed of the $E$-equation, $k_{h}$ -- 1 $h$-evolution equations and $k_{l}$ $l$-evolution equations system.  
  
The expression (\ref{plain-tem}) for emergent JBB time candidate is now (with the $\underline{B}$'s and extremization thereover {\sl absent} in the eliminated case), 
\be
\lt^{\se\sm(\sJ\sB\sB)} = \stackrel{\mbox{\scriptsize extremum $\d\underline{B}$ of Rot($d$)}}
                                                            {\mbox{\scriptsize of $S_{\mbox{\tiny JBB}}^{\mbox{\tiny RPM}}$}}
\left(\int\sqrt{           \{   \d h^2 +  h^2||\d_{\underline{B}} \mbox{\boldmath$l$}||_{\mbox{\boldmath\scriptsize$M$}_l}\mbox{}^2  \}/
                       {    2\{E_{\sU\sn\si} - V_{h} - V_{l} - J\}  }            } \right) 
\mbox{ } .  \label{TorreBruno}
\eeq
Note that such an absense also occurs in the GR case. 
This is via $h_{\mu\nu} = a^2 u_{\mu\nu}$ leading to $\{\d - \pounds_{\d{F}}\}\{a^2u_{\mu\nu}\} = a^2\{\{{\d a}/{a}\}u_{\mu\nu} + 
\d u_{\mu\nu} - \mD\mbox{}_{(\mu}\d{F}_{\nu)} + 0\} = a^2\{\d - \pounds_{\d{F}}\}u_{\mu\nu}$. 
Here, $a$ is the scalefactor and $\mD_{\mu}$ is the covariant derivative associated with $u_{\mu\nu} = h^{1/3}h_{\mu\nu}$. 
The 0 here arises from the constancy in space of the scalefactor in the role of conformal-factor killing off the extra conformal connection.
By this observation, scale--shape split approximate JBB time (and the approximate WKB time which coincides with it) avoids having a Sandwich/Best Matching Problem. 


\subsection{$h$ = scale approximation}\label{iudex}

The $h$-approximation to the action (\ref{JHL2}) is\footnote{$E_h$ is only approximately equal to $E_{\tU\tn\ti}$ since the $h$ and $l$ subsystems can interchange energy.}  
$\fS^{\sR\sP\sM}_{\sJ\sB\sB(h)} = \sqrt{2}\int \sqrt{\{E_{h} - V_{h}\}}\d h$.
Then the conjugate momenta are $P^{h} = \Last^{h} h$, the quadratic energy constraint is $\scE_{h}:= P^{h\, 2}/2 + V_{h} = E_{h}$  and the evolution equations are 
$\Last^{h} P^{h}   =  -\pa V_{h}/\pa h^{\ip\mu}$. 
This assumes that (using the subscript j to denote `judging')  
\beq
\mbox{(ratio of force terms)} 
\mbox{ } , \mbox{ } \mbox{ } 
F_{\sj} := \{\pa J/\pa h\} \left/ \{\pa V_{h}/\pa h\} \right.  = \{\pa J/\pa \mS\} \left/ \{\pa V_{\sS}/\pa \mS\} \right.  
\mbox{ } , \mbox{ } \mbox{ } 
\mbox{is of magnitude } \mbox{ } \epsilon_{\sss\sd\sss-1\sj} <<  1 \mbox{ } ,  
\label{SSA2}
\eeq
\beq
\mbox{(ratio of geometrical terms)} 
\mbox{ } , \mbox{ } \mbox{ } 
G_{\sj} := h ||\d \bl||^2_{\sbM}/d^2 h = \rho||\d \bS||^2_{\sbM}/\d^2 \rho  
\mbox{ } , \mbox{ } \mbox{ } 
\mbox{is of magnitude } \mbox{ } \epsilon_{\sss\sd\sss-2\sj} <<  1 \mbox{ } .
\label{SSA3}
\eeq
I originally considered an action-level scale-dominates shape' approximation \cite{Cones}, that is most clearly formulated as 
\beq
F := J/W_h = J/W_{\rho} 
\mbox{ } , \mbox{ } \mbox{ } 
\mbox{is of magnitude } \mbox{ } 
\epsilon_{\sss\sd\sss-1} << 1 \mbox{ } , 
\label{sds-1}
\eeq
\beq
G := ||\d_{\underline{B}}\bl||_{\sbM}/\d\{ \mbox{ln} \, h\} = ||\d_{\underline{B}}\bS||_{\sbM}/\d\{ \mbox{ln} \, \rho\} 
\mbox{ } , \mbox{ } \mbox{ } 
\mbox{is of magnitude } \mbox{ } 
\epsilon_{\sss\sd\sss-2} << 1 \mbox{ } .
\eeq
Each pair 1 and 1j, and 2 and 2j, are dimensionally the same but differ in further detail.
However, further consideration (Sec \ref{+temJBB}) reveals that this assumption is better justified if done by judging at the level of the equations of motion/forces. 
An example of this is how the effect of Andromeda on the solar system is not negligible at the level of the potential, but it is at the level of the tidal forces 
[which have an extra two powers of 1/(distance to Andromeda)].
Thus one is to use (\ref{SSA2}, \ref{SSA3}).


Then (\ref{yet-another-E}) can be taken as an equation for $t^{\se\sm(\sJ\sB\sB)}$ via the momentum--velocity relation [and this follows suit in the multi-$h$ case]. 
$\Last^{h} := \pa/\pa t^{\se\sm(\sJ\sB\sB)}_{h}$, now corresponding to [c.f. (\ref{TorreBruno})].
The approximate emergent JBB time candidate is then
\be
\lt^{\se\sm(\sJ\sB\sB)}_{h} = 
\left.
\int   \d h_{(0)} 
\right/
\sqrt{2\{E_{h} - V_{h_{(0)}}\}}  \mbox{ } ,
\label{hint2} 
\ee
which is of the general form 
\beq
\lt^{\se\sm(\sJ\sB\sB)}_h = {\cal F}[h, \d h] \mbox{ } . 
\label{hdh}
\eeq
N.B. that for this split and to this level of approximation, there is no $\fG$-correction to be done. 
This is because the rotations act solely on the shapes and not on the scale; in other words Configurational Relationalism is trivial here.

Finally, the first approximation to the $l$-equations is 
\beq
P^{l}_{\sfa}        =  h^2 \ttM_{\sfa\sfb} \Last^{h}l^{\sfb}
\Last^h P_{\sfa}^{l}   =  h^2 \ttM_{\sfa\sfb,\sfc} \Last^h l^{\sfb}\Last^h l^{\sfc} - \pa\{V_{l} + J\}/\pa l^{\sfa}  
\mbox{ } , 
\eeq
with the same notational interpretation as in Sec \ref{RPM-H-L-2} ($\ttM$ is the shape space metric). 
See \cite{FileR} for extension to the case of multiple $h$ degrees of freedom. 
Whenever we get disagreement with experiment, going back to the first, chroniferous formulation should be perceived as a possible option. 
Early 20th century `lunar anomalies' are an archetype for this. 
This is as per de Sitter's comment \cite{deSitter} {\it ``the `astronomical time', given by the Earth's rotation. 
Furthermore it was used in all practical astronomical computations.
It differs from the `uniform' or `Newtonian' time, which is defined as the independent variable of the equations of celestial mechanics."}

\subsection{Expansion of the isolated emergent-time equation}

\noindent  Pure-$h$ expressions of the general form (\ref{hdh}) are unsatisfactory from a Machian perspective. 
This is because since they do not give $l$-change an opportunity to contribute to the timestandard.  
This deficiency is to be resolved by treating them as zeroth-order approximations in an expansion involving the $l$-physics too.
Expanding (\ref{TorreBruno}), one obtains an expression of the form 
\beq
\lt^{\se\sm(\sJ\sB\sB)}_{(1)} = {\cal F}[h, l, \d h, \d l] \mbox{ } .  
\label{callie}
\eeq
[We also now write $t_0^{\se\sm(\sJ\sB\sB)}$ in place $t_h^{\se\sm(\sJ\sB\sB)}$.  
More specifically, for $h = \rho$ and $l^{\sfa} = \mS^{\sfa}$, 
\beq
\lt^{\se\sm(\sJ\sB\sB)}_{(1)} = \lt^{\se\sm(\sJ\sB\sB)}_{(0)} + 2^{-3/2}\int {\d \rho}\{F + G^2\}/{\sqrt{W_{\rho}}} + O(F^2) + O(G^4) \mbox{ } .
\label{Cl-Expansion}
\eeq
Thus one has an interaction term and an $l$-change term. 
Moreover the negligibility of $O(F^2)$ is controlled by judging criterion $F_{\sj}$ and that of $O(G^4)$ by $G^2_{\sj}$.  
This analysis is limited by how the correction terms are themselves determined by solving further equations, so that the emergent-time equation is part of a coupled system.
However the general form (\ref{callie}) itself is unaffected by this coupled nature.
The perturbative scheme of Sec 2.7 is a simple example of taking this further feature into account.

\subsection{First Approximation: Machian Classical Scheme}

The idea is then to perturbatively expand the energy equation, $l$-evolution equations and $h$-evolution equations. 
(The latter is purely an ancillary judging equation in the case of 1 $h$ degree of freedom.)   
For the energy equation to serve as a chroniferous equation, the analysis must be carried out at in $Q^{\sfA}$, d$Q^{\sfA}$ variables.
Contrast with how most classical and essentially all quantum perturbation theory are carried out in $Q^{\sfA}, P_{\sfA}$ variables.

What is being developed here is a Semiclassical Quantum Cosmology analogue of the astronomers' ephemeris time procedure. 
Suppose that one has found an accurate enough time for one's purposes. 
Then one can indeed revert to an analysis in terms of $Q^{\sfA}, P_{\sfA}$ variables for features within that universe that are fine enough to not contribute relevant change to 
the timestandard. 
This is very much expected to cover all uses of QM perturbation theory that apply to modelling laboratory experiments.
Here, fairly large-scale features of the Universe are expected to contribute a small amount in addition to the zeroth-order expansion of the universe and homogeneous-matter-mode contributions.  
There is a limit on ephemeris time schemes, since the iterations in those were at a level of form-fitting rather than a perturbative expansion of the equations of motion themselves.  
This is as opposed to just the specific situation of finding accurate timestandards on Earth necessitates more general analysis.
However we do not discard the possibility of being able to do form-fitting for the practical cosmological situation 
of an approximately-FLRW universe that models our own observed universe.
 
Rather than simply having a $t$ and using it as independent dynamical variable, 
one first considers time to be a highly-dependent variable until one has a satisfactorily accurate notion of time. 
Only then does one reinterpret this time as a convenient independent dynamical variable.

Note that gravitational Solar-System and cosmological GLET's could in principle differ from atomic clock readouts.  
However, to date there is no evidence of any such discrepancy. 
This was an important check in proposing atomic clocks in the first place \cite{Parry}.

\subsection{First assessment of Semiclassical Quantum Cosmology's approximations}

This reveals difficulties wigth some details the quantum cosmological status quo (Sec 3) by comparison with the conventional practise in the far more carefully studied and 
experimentally tested arena of classical dynamics.
These discrepancies do {\sl not} concern the Machianization of Semiclassical Quantum Cosmology itself. 
(This is, rather, a constructive import from Dynamics and Astronomy to {\sl whatever form Semiclassical Quantum Cosmology should take}). 
Rather some of the plethora of approximations conventionally made to simplify the Semiclassical Quantum Cosmology equations.  

{            \begin{figure}[ht]
\centering
\includegraphics[width=0.42\textwidth]{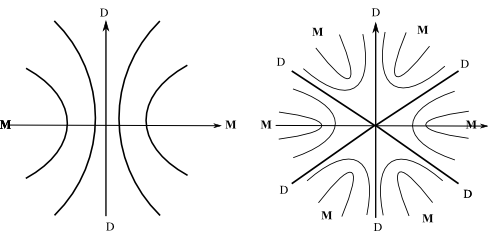}
\caption[Text der im Bilderverzeichnis auftaucht]{        \footnotesize{Contours on configuration space for single and triple negative power potentials 
(the 1-$d$ 3-particle case for simplicity). 
These have abysses along the corresponding double collision lines D and high ground in between these. 
(For negative-power coefficients such as for the attractive Newtonian Gravity potential.) 
M are the merger configurations (with the third particle at the centre of mass of the other two).  
}        }
\label{Fig--5}
\end{figure}  }

\noindent Classical Problem 1) Consider e.g. Newtonian Gravity/RPM's that model dust-filled GR cosmology.
%
%
Then on the corresponding lines of double collision, D, the potential has infinite abysses/peaks (Fig \ref{Fig--5}). 
\noindent The `scale dominates shape' approximation is thus certainly not valid near there, so some assumptions behind the Semiclassical Approach fail in the region around these lines.  
Thus for negative powers of relative separations the heavy approximation only makes sense in certain wedges of angle. 
There is then the possibility that dynamics set up to originally run in such regions falls out from them. 
I.e. a stability analysis is needed to determine whether semiclassicality is representative. 
I.e. there is a tension between the procedure used in Semiclassical Quantum Cosmology 
and  the futility of {\sl trying to approximate a 3-body problem by a 2-body one} \cite{ScaleQM, FileR}.  

\noindent Classical Problem 2) Semiclassical Quantum Cosmology's neglect of the $T_l$ term is part of decoupling the $h$ and $l$ subsystems, 
which contributes to rendering them easier to solve analytically. 
However, the classical dynamics version of this (shape--scale split 1- or 2-$d$ RPM version) involves {\sl throwing away the central term}. 
I.e. the mathematical equivalent of neglecting the centrifugal barrier in the study of planetary motion. 
This causes unacceptable quantitative and qualitative errors (linear motion versus periodic motion in the shape of an ellipse).
%
%
%
This qualitative difference indeed carries over to the RPM counterpart \cite{FileR}.  
%
%

\subsection{Modelling assumptions for the perturbative Classical Machian Scheme}\label{AppA}

\noindent 1) A key feature is that what is conventionally an `independent variable' $t$, is here a quantity emergent from change, 
and with all change having the opportunity to contribute.  
I.e. time is here ab initio a {\sl highly dependent} variable.  
Thus it is clear that $t^{\se\sm(\sJ\sB\sB)}$ itself is to be subjected to perturbations, whereas the conventional $t$ itself is not.  

\noindent 2) Due to the way approximations are to be judged, we want to keep $\pa_{l}V_{l}$ but not $V_{l}$, and we want to judge $J$ partly via $\pa_{h}J$.

\noindent 3) We also need a vector of $\epsilon$'s, $\underline{\epsilon}$, rather than a single small parameter, since we have multiple a-priori independently small quantities.
This is more usual in Theoretical Physics than 1) or 2). 
E.g. the $\lambda \varsigma^3 + \mu \varsigma^4$ interaction potential in the QFT of a scalar field, $\varsigma$.
The general case will become further complicated if some $\epsilon$'s can be small that they are around the size of nontrivial powers of other $\epsilon$'s.
Three regimes of particular tractability are i) $\underline{\epsilon} = (0, ..., 0, \epsilon, 0, ... 0)$: approximation by a single $\epsilon$. 
ii) $\underline{\epsilon} = (\epsilon, \epsilon, ... \epsilon)$: `$\epsilon$-democracy' in which all $\epsilon$'s are {\sl roughly} the same size.  
iii) The partial vector with a single $\epsilon$ on $> 1$ entries and the other entries zero.  
Note that a given $\epsilon$ can be forced to be big by circumstance. 
Then one has a perturbation scheme with one $\epsilon$ less, though the awareness and formalism remain similar.  
Some papers \cite{KS91, BK97, Kiefersugy} investigate Quantum Cosmology by expanding in a single parameter.  
The present Article and \cite{FileR} then systematize the treatment of this.


All in all, we take 
\beq
Q^{\sfA} = Q^{\sfA}_{(0)} + \underline{\epsilon}\cdot\underline{Q}^{\sfA}_{(1)} + O(\epsilon^2) \mbox{ } , 
\label{A-1}
\eeq
\beq
t^{\se\sm} = t^{\se\sm}_{(0)} + \underline{\epsilon}\cdot\underline{t}^{\se\sm}_{(1)} + O(\epsilon^2) \mbox{ } ,
\label{A-2}
\eeq
though the $\underline{\epsilon}$ is taken to originate from the primed expressions for the energy equation (\ref{E-prime}) and classical $l$-equations of motion (\ref{l-EOM-prime}).  
Then one's classical system is (the $\Gamma$ is a Christoffel symbol)
\beq
h^{\SStar\SStar} = \epsilon_{2\sj} h^2 M_{\sfb\sfc} l^{\sfb \SStar} l^{\sfc \SStar} - {\pa V_{h}}/{\pa h} - \epsilon_{1\sj} {\pa J^{\prime}}/{\pa h}  \mbox{ } , 
\eeq
\beq
h^{\SStar\, 2} + \epsilon_2 h^2 M^{\prime}_{\sfb\sfc} l^{\sfb \SStar} l^{\sfc \SStar} = 2\{E_{\sU\sn\si} - V_{h} - \epsilon_3 V_{l}^{\prime} - \epsilon_1 J^{\prime}\}  \mbox{ } ,
\label{E-prime}
\eeq
\beq
l^{\sfa \SStar\SStar} + {\Gamma^{\sfa}}_{\sfb\sfc} l^{\sfb \SStar} l^{\sfc \SStar} + {2 h^{\SStar} l^{\sfa \SStar}}/{h} = - h^{-2}N^{\sfa\sfb}
\left\{
{\pa V_{l}}/{\pa l^{\sfb}} + \epsilon_4 {\pa J^{\prime}}/{\pa l^{\sfb}}
\right\} \mbox{ } .  
\label{l-EOM-prime}
\eeq
Thus we have a string of six $\epsilon$'s.
\noindent Note that in the absence of $V_{l}$, using an $\epsilon_4$ may be undesirable.
For more than 1 $h$ degree of freedom, $\epsilon_{2\sj}$ and $\epsilon_{1\sj}$ are not just judging equations but also enter the system of equations to be solved. 
Also, we {\sl decide} to take $h$ as heavy on the basis of the size of $\epsilon_{1\sj}$, it is $\epsilon_3$ itself that enters the subsequent working.  
Thus classically we have a $\underline{\epsilon}$ with 4 components: small $l$-kinetic term, small $l$-potential term, 
small interaction potential and small interaction force on the $l$-system.  
Finally, note that $\epsilon_3$ is $l$ subdominance to $h$ in the potential, i.e. $|V_{l}/V_{h}|$ small, and $\epsilon_{4}$ is $|\pa J/\pa l \, / \, \pa V_{l}/\pa l|$ small.  

\mbox{ } 

\noindent {\bf Zeroth-order classical Machian equations.}
%
Applying (\ref{A-1}) for $Q^{\sfA}$ = $h$, $l^{\sfc}$ and (\ref{A-2}) and Taylor-expand $M$, $N$, $\Gamma$, $V_{h}$, $V_{l}$, $J$ and their derivatives
gives back to zeroth order the expected equations,  
\beq
h_{(0)}^{\SStar_{(0)}\SStar_{(0)}} =  - \pa V_{h}(h_{(0)})/\pa h_{(0)} \mbox{ } , 
\eeq
\beq
h_{(0)}^{\SStar_{(0)} 2} = 2\{E_h - V_{h}(h_{(0)})\}  \mbox{ } ,
\label{E-zero}
\eeq
\beq
l_{(0)}^{\sfa \SStar_{(0)}\SStar_{(0)}} + {\Gamma^{\sfa}}_{\sfb\sfc}\big(l_{(0)}^{\sfp}\big)l_{(0)}^{\sfb \SStar_{(0)}}l_{(0)}^{\sfc \SStar_{(0)}} + 
{2h_{(0)}^{\SStar_{(0)}}l_{(0)}^{\sfa \SStar_{(0)}}}/{h_{(0)}} = - h_{(0)}^{-2}{N^{\sfa\sfb}\big(l^{\sfp}_{(0)}\big)}{\pa V_{l}(l^{\sfp}_{(0)})}/{\pa l^{\sfb}_{(0)}} 
\mbox{ } .  
\label{l-EOM-zero}
\eeq
\noindent {\bf First-order classical Machian equations}.
%
These are 
%
%
\beq
\underline{\epsilon} \cdot 
\left\{  
h_{(0)}^{\SStar_{(0)}} \{ \underline{h}_{(1)} - 
                            \underline{t}_{(1)}^{\SStar_{(0)}}h_{(0)}^{\SStar_{(0)}} \} + 
\underline{h}_{(1)} {\pa V_{h}(h_{(0)})}/{\pa h_{(0)}}
\right\}
+ {\epsilon_{2}}h_{(0)}^2 M_{\sfc\sfd}^{\prime}\big(l_{(0)}^{\sfp}\big)l_{(0)}^{\sfc \SStar_{(0)}}l_{(0)}^{\sfd \SStar_{(0)}}/2 
+ \epsilon_3V_{l}^{\prime}\big(l_{(0)}^{\sfc}\big) + \epsilon_1 J^{\prime}\big(h, l_{(0)}^{\sfc}\big) \mbox{ } ,   
\label{E-1}
\eeq
$$
\underline{\epsilon}\cdot
\left\{
\underline{l}_{(1)}^{\sfb \SStar_{(0)} \SStar_{(0)}} - \underline{t}_{(1)}^{\SStar_{(0)} \SStar_{(0)}}l_{(0)}^{\sfb \SStar_{(0)}} - 
2 l_{(0)}^{\sfb\SStar_{(0)}\SStar_{(0)}}\underline{t}_{(1)}^{\SStar_{(0)}} + 
\underline{l}_{(1)}^{\sfp} \frac{\pa \Gamma^{\sfb}_{\sfa\sfc}\big(l_{(0)}^{\sfd}\big)}{\pa l^{\sfp}_{(0)}}
l_{(0)}^{\sfa \SStar_{(0)}}l_{(0)}^{\sfc \SStar_{(0)}} + 2\Gamma^{\sfb}_{\sfa\sfc}\big(l_{(0)}^{\sfd}\big) l_{(0)}^{\sfc \SStar_{(0)}}
\{\underline{l}_{(1)}^{\sfa \SStar_{(0)}} - \underline{t}_{(1)}^{\SStar_{(0)}}l_{(0)}^{\sfa \SStar_{(0)}}\} 
\right.
$$
$$
+ \frac{2}{h_{(0)}} 
\left\{
\underline{l}_{(1)}^{\sfb \SStar_{(0)}}h_{(0)}^{\SStar_{(0)}}  + l_{(0)}^{\sfb \SStar_{(0)}}\underline{h}_{(1)}^{\SStar_{(0)}} - l_{(0)}^{\sfb \SStar_{(0)}}h_{(0)}^{\SStar_{(0)}}
\left\{  
\frac{\underline{h}_{(1)}}{h_{(0)}} + 2\underline{t}_{(1)}
\right\}   
\right\}
\left. 
- \frac{N^{\sfa\sfb}\big(l_{(0)}^{\sfq}\big)}{h_{(0)}^2} 
\left\{ 
\frac{2\underline{h}_{(1)}}{h_{(0)}}\frac{\pa V_{l}(l_{(0)}^{\sfq})}{\pa l^{\sfa}_{(0)}}  - 
\underline{l}_{(1)}^{\sfp} \frac{\pa^2V_{l}(l^{\sfq}_{(0)})}{\pa l^{\sfa}_{(0)}\pa l^{\sfp}_{(0)}}
\right\}
\right.
$$
\beq
\left.
+ \frac{\underline{l}_{(1)}^{\sfc}}{h_{(0)}^2}  \frac{\pa N^{\sfb\sfa}\big(l^{\sfq}_{(0)}\big)}{\pa l^{\sfc}_{(0)}}\frac{\pa V_{l}\big(l^{\sfq}_{(0)}\big)}{\pa l^{\sfa}_{(0)}}
\right\}  
= - \epsilon_3 \frac{N^{\sfa\sfb}\big(l^{\sfq}_{(0)}\big)}{h_{(0)}} \frac{\pa J(h_{(0)}, l^{\sfq}_{(0)})}{\pa l^{\sfa}_{(0)}}
\label{l-EOM-1}  \mbox{ } .  
\eeq
Note that one cannot just cancel the $\epsilon$'s off in general case (unlike for schemes with just the one $\epsilon$).  
Also the system given is indeed well-determined.
One can take the quantities to be solved for at each step to be as follows. 
\noindent $h(t^{\se\sm(\sJ\sB\sB)})$ = $I$($t^{\se\sm(\sJ\sB\sB)}$), $l^{\sfa}$($t^{\se\sm(\sJ\sB\sB)}$) = S$^{\sfa}$($t^{\se\sm(\sJ\sB\sB)}$)
\noindent Then $t^{\se\sm(\sJ\sB\sB)}$($h$, $l^{\sfa}$) = $t^{\se\sm(\sJ\sB\sB)}$($I$, S$^{\sfa}$).  
\noindent This transcends to the GR case.  
\noindent It can be investigated perturbatively.
\noindent $\underline{t}^{\se\sm(\sJ\sB\sB)}_{(1)}$ is not a separate entity but rather abstracted from $\underline{h}_{(1)}$ and $t^{\se\sm(\sJ\sB\sB)}_{(0)}$ (or $h_{(0)}$).

\subsection{Concrete RPM example of Machian Classical Scheme: 3-stop metroland example}

In this case, action (\ref{JHL2}) can further be expressed as  
\beq
S_{\sr\se\sll} = \sqrt{2}\int \sqrt{\d\rho^2 + \rho^2 \d\varphi^2}\sqrt{E_{\sU\sn\si} - A\rho^2 - B\rho^2\mbox{}\mbox{cos}\,2\theta}
\eeq
This is for an HO potential which takes the given form once expressed in the scale--shape coordinates $\rho$, $\varphi$.  
Set I = $h$ and $\varphi$ = $l$. 
Note that this example simplifies by having no $V_{l} = V_{\theta}$ and hence no $\epsilon_2$.

\noindent The $\epsilon$-scheme is now:
\beq
h^{\SStar\SStar} = \epsilon_{2\sj} h\{\theta^{\SStar\,2}/\epsilon_{2\sj}\} + 2h\{A + \epsilon_{1\sj} B^{\prime}\mbox{cos}\,2l\} \mbox{ } ,  
\label{h-prime-3}
\eeq
\beq
h^{\SStar\, 2} + \epsilon_2\{h^2\theta^{\SStar\,2}/\epsilon_2\} = 2\{E_{\sU\sn\si} - Ah^2 - \epsilon_1 B^{\prime}h^2\mbox{cos}\,2l\}  \mbox{ } ,
\label{E-prime-3}
\eeq
\beq
l^{\SStar\SStar} + 2{h^{\SStar}\theta^{\SStar}}/{h}  = - 2\epsilon_3\,B^{\prime\prime}\,\mbox{sin}\,2l \mbox{ } .  
\label{l-prime-3}
\eeq
\mbox{ } \mbox{ } The zeroth order then gives back the judging equation 
$
h_{(0)}^{\SStar_{(0)}\SStar_{(0)}} = 2A\,h_{(0)}
$
and the system 
\beq
h_{(0)}^{\SStar_{(0)}\, 2}  = 2\{E_h - A\,h_{(0)}^2\}  \mbox{ } , \mbox{ } \mbox{ }
h^{(0)\, 2}\theta^{\SStar} =  \fD  \mbox{ } . 
\label{E-zero-3}
\eeq
Here, a first integral has been performed on the last equation. 
$\fD$ is a conserved relative dilational momentum quantity \cite{FileR}.  


This example serves to illustrate the aforementioned  with neglecting the $l$-kinetic term.  
Had that been kept, the second equation of (\ref{E-zero-3}) can then be used to provide an $h$-equation of a qualitatively distinct form, 
\beq
h_{(0)}^{\SStar\, 2} + \fD^2/h_{(0)}^2 = 2\{E_h - Ah_{(0)}^2\}  
\eeq
(see below example for the significance of this).  
Also then 
\beq
l_{(0)}^{\SStar_{(0)}\SStar_{(0)}} + 2{h_{(0)}^{\SStar_{(0)}}l^{\SStar_{(0)}}}/{h_{(0)}}  = -2B\,\mbox{sin}\,2l_{(0)} \mbox{ } .
\eeq
Note 1) The effect of inclusion on this example is not the `1/$r$' potential case's Keplerian ellipses versus straight lines. 
However, the difference between keeping the central term or not is qualitatively significant over the whole set of central force problems rather than just the $1/r$ potential case. 
For the present example's HO's, this is the difference between ellipses centred on the origin (exact):
$
h_{(0)} = 1/\sqrt{\Biggamma + \Bigeta\,\mbox{cos}^2(\varphi_{} - \varphi_{0})} 
$
, and spirals (inexact):
$
h_{(0)} = \Bigbeta/\sqrt{1 + \Bigalpha\{\varphi - \varphi_0\}}
$
for constants $\Bigalpha, \Bigbeta, \Biggamma, \Bigeta$.  
%
%
A forteriori, at the qualitative level, the ellipses are periodic and the spirals are not.    

One might also choose not to regard $\epsilon_1$ as small here so as to support nontrivial physics to first order.  
Then the last equation has a $-2B\,\mbox{sin}\,2l$ right-hand side.
This can be thought of in terms of $V_{l} = 0$ makes lowest order $l$-dynamics trivial.  
Moreover, this is a common feature for scaled RPM's.  
Thus taking only the lowest iteration in the semiclassical approach for these models can be of limited applicability, as it may not capture any $l$-dynamics at all.  
\cite{SemiclIII} already went further than that, but under more restrictive assumptions on what is perturbed that themselves lack in Machianity.
(If $t$ is perturbed, so should the $Q^{\sfA}$ from whose change d$Q^{\sfA}$ the Machian time is to be abstracted...).  

Note that $J$ and $J_{,l}$ sometimes have the same form here. 
(For these trig functions, $l$ is near $\pi/4$), thus illustrating that $\epsilon_{1\sj}$ can cease to be a separate diagnostic.
This teaches us that such schemes really only work out for certain regions (i.e. are local and thus non-global.)  
$|B| << A$ helps ensure some $\epsilon$'s are suitably small, but other conditions favour mostly-radial motion.
(I.e. scale dominates shape, so this is also in accord with cosmological modelling.) 
This is alongside confinement of these wedges to suitably small values of cos$\,\varphi$ and sin$\,\varphi$.

\subsection{Other classical Problem of Time facets for classical RPM's}\label{Others}

For the present paper, the beables are of the form \cite{AHall} $\mbox{K = {\cal F}[$\rho, S^{\sfA}, p_{\rho}, P_S^{\sfA}$ alone]}$. 
A fourth Problem of Time facet is the Constraint Closure Problem.  
However, for RPM's evaluating the Poisson brackets of the constraints readily demonstrates that this is classical absent \cite{FileR}.
\noindent These Mechanics models do not have a GR-like spacetime structure. 
Thus the Spacetime Reconstruction Problem and Foliation Dependence Problem of Time facets are non-issues for RPM's. 
\noindent A local resolution of the Problem of Time being as much as is attempted, that is all of the facets to overcome for now.
(The other facets this precludes are the multiple-choice and global problems \cite{Kuchar92, I93, APoT2}.)

\section{Semiclassical Approach to RPM's}\label{Semicl}

\noindent This scheme starts from the $h$--$l$ split of Sec \ref{+temJBB}.
I choose to use reduced scale = $h$, shape = $l$ models for most detailed work.  
This work is Sec 2's successor in  including at the level of implementing Mach's Time Principle in the `GLET is to be abstracted from STLRC' way.  
The current Sec interpolates between classical and quantum forms of perturbation theory, which is nonstandard.

\subsection{QM of the r-formulation of RPM}

I use conformal operator ordering because it is configuration space coordinate invariant and preserves a straightforward invariance of the relational product-type action.
(This is where \cite{Banal} Misner's conformal invariance \cite{Magic} can most deeply and simply be traced to.)  
Then the classical energy equation's $N^{\sfA\sfB}p_{\sfA}p_{\sfB}$ becomes the conformal-ordered $\triangle^{\scc} := \triangle - c(N, d)$.  
Here $\triangle_{\sfS(N, d)}$ is the Laplacian on the corresponding shape space, and \cite{FileR}
\beq
c(N, d) := 0    \mbox{ for \mbox{ }                    $d$ = 1         } \mbox{ and } \mbox{ }                   
                                    {        {3 n\{2 n - 3\}}/{4\{n - 1\}} }   {\mbox{ for        $d$ = 2                      }}
\eeq
for the RPM's covered in the current article.


Then the solvable scaled-RPM series' TISE (time-independent Schr\"{o}dinger equation is \cite{QuadIII}  
\beq
-\hbar^2\{\pa^2_{\rho} + q(N, d)\rho^{-1}\pa_{\rho} - \rho^{-2}c(N, d) + \triangle_{\sfS(N, d)}\}\Psi = 2\{E_{\sU\sn\si} - V(\rho, S^{\sfa})\}\Psi
\label{Gilthoniel}
\eeq
for particle number $N \geq 3$ for relational nontriviality \cite{FileR} and dimension $d$ = 1 ($N$-stop metrolands) or 2 ($N$-a-gonlands).
\noindent For later use, let $\fS(N, d)$ denote the shape space (scale-free relational configuration space) for $N$-particle $d$-dimensional RPM, 
$
\fS(N, d) = \fQ(N, d)/\mbox{Sim}(d)
$
for $\fG =$ Sim($d$) the $d$-dimensional similarity group of translations, rotations and dilations.  
$
\mbox{dim(Sim($d$))} = \mbox{dim(Tr($d$))} + \mbox{dim(Rot($d$))} + \mbox{dim(Dil($d$))} = d + d\{d - 1\}/2 + 1 = d\{d + 1\}/2 + 1.  
$
and
$ 
q(N, d) := \mbox{dim}(\fS(N, d)) = \mbox{dim}(\mathbb{R}^{Nd}/\mbox{Sim}(d)) = Nd - \{d\{d + 1\}/2 + 1\} = nd - 1 - d\{d - 1\}/2.  
$

(\ref{Gilthoniel}) then separates into scale and shape parts. 
The scale part solved for the general free and isotropic-HO potential cases \cite{QuadIII}. 
It is solved in various cases in \cite{FileR, MacFarlane, QuadIII}.

\subsection{Born--Oppenheimer (BO) scheme and its quantum-cosmological analogue} 

I take this first step of the Semiclassical Approach to mean ansatz (\ref{BO}) alongside the following approximations.

\subsubsection{BO approximation} 

\noindent Let $\widehat{C} := \hat{H}  -  \hat{T}_{h}$: the complement of the heavy kinetic term.
The $|\chi\rangle$-wavefunction expectation value (integrated over the $l$ degrees of freedom, i.e. `$l$-averaged') is then 
$
c := \langle\chi|\widehat{C}|\chi\rangle = \int_{\sfS(N, d)}\chi^{*}(h, l^{\sfa}) \, \widehat{C}(h, l^{\sfa}, p^{l}_{\sfa}) \, \chi(h, l^{\sfa})\, \mathbb{D} l.  
$ 
The associated integration is over the $l$ degrees of freedom and so over RPM's shape space, with $\mathbb{D}l = \mathbb{D}\mS$ the measure over the shape space $\fS(N, d)$].  
%
%
The $|\chi\rangle$ sometimes requires suffixing by its quantum numbers, which I take to be multi-indexed by a single straight letter, k.  
Thus the above $c$ is, strictly, $c_{\sk\sk}$ and there is an obvious off-diagonal equivalent 
$
c_{\sk\sll} := \langle\chi_{\sk}|\hat{C}|\chi_{\sll}\rangle.  
$ 
{\it The BO approximation} alias `diagonal dominance condition' is then that 
$
\mbox{for} \mk \neq \ml, \mbox{ } \left|c_{{\sk}{\sll}}/c_{{\sk}{\sk}}  \right| =: \epsilon_{\sB\sO} << 1.    
$
Assuming that this holds, one then considers $\langle\chi| \times$ the TISE with the Born--Oppenheimer (BO) ansatz substituted in.

\subsubsection{Adiabatic approximations}

The $h$-derivatives acting upon the product ansatz wavefunction ansatz produce multiple terms by the product rule,   
%
$
|\chi\rangle \pa_{h}^2\psi
$,  
$
\pa_{h}\psi \pa_{h}  |\chi\rangle
$, 
$
\psi\pa^2_{h}  |\chi\rangle  
$.  
The first term is always kept.  
BO themselves discarded the next two for being far smaller than the first (a first kind of {\sl adiabatic approximation}: $h$-changes in $\chi$ are much smaller than those in $\psi$).   
However (Sec \ref{VaNiel}) the emergent semiclassical time approach to the Problem of Time requires keeping at least one such cross-term. 
This is a case of qualitative importance overriding smallness. 
In the second term of (\ref{Gilthoniel}), $|\chi\rangle \pa_h \psi$ is kept and $\psi \pa_h |\chi\rangle$ is usually discarded.

\subsubsection{Commentary on adiabatic-type terms}

As well as the types of adiabatic terms already present at the classical level as covered in Sec \ref{+temJBB}, 
there are two different `pure forms' that adiabaticity can take at the quantum level.  
`a($l$)': quantities that are small through $|\chi\rangle$ are far less sensitive to changes in $h$- rather than $l$-physics (the $l$ stands for `internal to the $l$-subsystem').  
`a($m$)': quantities that are small through $|\chi\rangle$ being far less sensitive to changes in $l$-physics than $\psi$ is to changes in $h$-physics 
(the $m$ stands for `mutual between the $h$ and $l$ subsystems').

Note that none of the above follow from the smallness of the classical adiabatic parameter $\omega_h/\omega_l$.
This is because some wavefunctions can be very steep or wiggly even for slow processes, e.g. the 1000th Hermite function for the slower oscillator.  
However, high wiggliness is related to high occupation number.  
This is via quantum states increasing in number of nodes as one increases the corresponding quantum numbers. 
Additionally, high occupation number is itself a characterization of semiclassicality.   
Also inspection of the $h$ and $l$ equations reveals that both a($l$) and a($m$) occur in terms also containing an $\epsilon_{\sss\sd\sss-1}$ as per (\ref{sds-1}).  
Thus, overall, these terms in the equations are particularly small.\footnote{There are in total 16 terms that are often neglected in the reduced RPM semiclassical system \cite{SemiclI}.
$h$ and $l$ in the enumeration denote which equation these terms feature in. 
\beq
\frac{\pa l}{\pa t} {\frac{\pa |\chi\rangle}{\pa l}}\left/{\frac{\pa |\chi\rangle}{\pa h}}\right. = \frac{\ma(l)}{\ma} := \epsilon_i\mbox{ } ,
\eeq 
i.e. a ratio of quantum $l$-subsystem adiabaticity to classical mutual adiabaticity.} 
Finally, note that Massar and Parentani's work on inclusion of non-adiabatic effects \cite{MP98} in the minisuperspace arena shares the present paper's spirit of 
considering qualitative effects of keeping usually neglected terms in the Quantum Cosmology equations.  
In this model, the effects found are i) couplings between expanding and contracting universes. 
ii) A quantum-cosmological case of the {\it Klein paradox} (backward-travelling wave generation from an initially forward-travelling wave).

\subsection{The WKB scheme}

I take this to consist of the subsequent ansatz \label{WKB-An} for the $h$-wavefunction alongside the following approximations.
%
%
For physical interpretation, I rewrite the principal function $S$ by isolating a heavy mass $M$, $S(h) = M F(h)$. 
[For 1 $h$ degree of freedom, this is trivial; for more than 1, it still makes sense if the sharply-peaked mass hierarchy condition holds.]
The associated {\it WKB approximation} is the negligibility of second derivatives, 
\be
\left|\frac{\hbar}{M}\frac{\pa_{h}^2F}{|\pa_{h}F|^2}\right| << 1 \mbox{ } . 
\ee
The associated dimensional analysis expression is $\hbar/MF =: \epsilon_{\sW\sK\sB^{\prime}} << 1$. 
This is to be interpreted as (quantum of action) $<<$ (classical action) via the reinterpretability of $S$ as classical action (see e.g. \cite{Lanczos}), 
which has clear semiclassical connotations.  


\noindent A further incentive for using 1 $h$ degree of freedom is that this trivially gets round having to explicitly solve nonseparable Hamilton--Jacobi equations.
This practical problem generally plagues the case of $> 1$ $h$ degrees of freedom \cite{Goldstein-BenFra}.

\subsection{The BO--WKB scheme's scale--shape split RPM $h$- and $l$-equations}

\noindent Then the r-formulation for RPM's {\sl $h$-equation}, $\langle\chi| \times$ [TISE (\ref{Gilthoniel})], with the BO and WKB ans\"{a}tze substituted in, is 
$$ 
\{\pa_{h}S\}^2 - i\hbar \, \pa_{h}\mbox{}^2S - 2i\hbar \, \pa_{h}S\langle\chi|\pa_{h}|\chi\rangle - \hbar^2\big\{  \langle\chi|  \pa_{h}\mbox{}^2  |\chi\rangle + 
k(N, d)h^{-1}  \langle\chi|\pa_{h}|\chi\rangle   \big\} - i\hbar h^{-1}k(N, d)\pa_{h}S 
$$
\beq
+ \hbar^2 h^{-2}\{c(N, d) - \langle\chi|\triangle_{l}|\chi\rangle\} + 2 V_{h}(h) + 2\langle\chi| V_l(l^{\sfa}) + J(h, l^{\sfa})|\chi\rangle = 2 E_{\sU\sn\si} \mbox{ } \mbox{ } . 
\eeq
\noindent Also, the r-formulation for the RPM is \{1 -- $\mP_{\chi}$\}[TISE (\ref{Gilthoniel})), and takes for now the fluctuation equation form 
\beq
\{1 - \mP_{\chi}\}  \big\{ - 2i\hbar \, \pa_{h}  |\chi\rangle  \pa_{h} S - \hbar^2\big\{  \pa_{h}\mbox{}^2  |\chi\rangle + k(N, d)h^{-1}  \pa_{h}  |\chi\rangle  + 
h^{-2}\triangle_{l}\}  |\chi\rangle + 2\{V_{l}(l^{\sfa}) + J(h, l^{\sfa})\}   |\chi\rangle \big\}  = 0  \mbox{ } .  
\label{l-TDSE-prime}
\eeq
Here, $\mP_{\chi}$ is the projection operator $|\chi\rangle\langle\chi|$.  
These equations were first given in \cite{QuadIII}.
Via various extensions in e.g. \cite{Datta97, Kieferbook, SemiclIII}, they generalize the equation given by Banks \cite{Banks}.

\subsection{Emergent WKB time}

It is then standard in the semiclassical approach to use that $\pa_{h}\mbox{}^2 S$ is negligible by the WKB approximation to remove the second term from the $h$-equation.
One then applies 
\beq
\pa_{h} S = p_{h} = \mbox{\Last}h \mbox{ } .
\label{lance2}
\eeq
Here $\Last := \d/\d t^{\se\sm(\sW\sK\sB)}$ is by identifying $S$ as Hamilton's function.  
Next, one uses the expression for momentum in the Hamilton--Jacobi formulation, the momentum-velocity relation, and the chain-rule to recast $\pa_{h}$ as $\pa_{h}t\, \Last$. 
%
%
\noindent Note that\footnote{This coincidence expands 
on comments by Barbour (personal communication), Kiefer \cite{Kiefer93} and Datta \cite{Datta97}.}
%
$t^{\se\sm(\sW\sK\sB)}_{(0)} = t^{\se\sm(\sJ\sB\sB)}_{(0)}$, so the notation can be simplified to $t^{\se\sm}_{(0)}$. 
It then follows from this identification and Sec \ref{+temJBB} that the approximate emergent WKB time is aligned with Newtonian time, proper time and cosmic time in various 
contexts but additionally now regarded as placed on a relational footing. 
Additionally, Sec \ref{+temJBB}'s properties and critiques extend to approximate emergent WKB time.  


The full (except for $\pa_{h}\mbox{}^2 S$ neglected) Machian $h$-equation is then  
$$ 
\{\Last h\}^2 - 2i\hbar  \Last h  \langle\chi|  \{\Diamond/\Last h\} \Last |\chi\rangle - 
\hbar^2 \{ \langle \chi | \Heart^2  | \chi \rangle + k(N, d) h^{-1} \langle \chi | \Heart | \chi \rangle \} 
$$
\beq
- i\hbar k(N, d) h^{-1} \Last h  +  \hbar^2 h^{-2} \{ k(\xi) - \langle \chi |  \triangle_{l}  | \chi \rangle \} = 
2\{  E_{\sU\sn\si} - V_{h}  -  \langle\chi|  V_{l}      |\chi\rangle  -  \langle\chi|  J   |\chi\rangle        \} \mbox{ } \mbox{ } .    
\label{TimeSet}
\eeq
Here, $\Diamond := \Last  - \Last l \, \pa_{l}$ and $\Heart := \Diamond/{\Last h}$ are the equation-simplifying groupings of derivatives.  
\cite{SemiclI} contained an antecedent of this, due to recognizing the derivative combination but only gave an example of the occurrence of such a term rather than the full $l$-equation.

Neglect the second, third, fourth, fifth, sixth and eighth terms and the $\Last l\,\pa_{l}$ contributions.
(See Secs \ref{Semi-Back}--\ref{Semi-Avs} for various possible justifications.)
Then this $h$-equation collapses to the standard semiclassical approach's Hamilton--Jacobi equation,    
\beq
\{\pa_{h}S\}^2 = 2\{E_h - V_{h}\} \mbox{ } , \mbox{ or } \mbox{ } 
\label{Sicut}
{\Last} h^2 = 2\{E_h - V_{h}\} \mbox{ } . 
\eeq
The second form is by (\ref{lance2}), and is especially justified because $S$ is a standard Hamilton--Jacobi function.


\noindent A reformulation of the latter that is of use in further discussions in this article is the analogue Friedmann equation, 
\beq
\{ \Last h/h\}^2 = 2E_{\sU\sn\si}/h^2 - 2V_{h}/h^{2} \hspace{0.3in} \mbox{($h$ = } \rho) \mbox{ } \mbox{ }   .
\label{Este}
\eeq                            
Whichever of the above forms is then solved by 
\beq
{\lt}^{\se\sm} = 2^{-1/2}          \int   {\d h}   \left/     {\sqrt{E_h - V_{h}}} \right.  \mbox{ } \mbox{ } .  
\label{Jesu}
\eeq

\subsection{Recasting of the $l$-fluctuation equation as a TDSE}\label{VaNiel}

\noindent One passes from a fluctuation equation to a semiclassical emergent time-dependent wave equation (TDWE) via the crucial chroniferous move 
\beq
N^{hh}  i\hbar  \frac{\pa S}{\pa h}               \frac{\pa \left| \chi\right \rangle}{\pa h} = 
i\hbar \, N^{hh}  p_{h}                                    \frac{\pa \left| \chi\right \rangle}{\pa h} =
i\hbar \, N^{hh}  M_{hh}  \Last h      \frac{\pa \left| \chi\right \rangle}{\pa h} = 
i\hbar \frac{  \pa h       }{  \pa t^{\se\sm(\sW\sK\sB)}  }\frac{\pa \left| \chi\right \rangle}{\pa h} \approx  
i\hbar                                                     \frac{\pa \left| \chi\right\rangle }{  \pa t^{\se\sm(\sW\sK\sB)}  }    \mbox{ } , 
\eeq
which proceeds via (\ref{lance2}) and the chain-rule in reverse.
\noindent In this paper's case, $N_{hh} = 1 = M^{hh}$; I include these, however, to show the greater generality of the working; in particular this is needed in GR examples. 

The full emergent semiclassical TDWE is then
\beq
i\hbar\{1 - \mP_{\chi}\} \Diamond |\chi\rangle = 
\{1 - \mP_{\chi}\}  \{ - \{\hbar^2/2\} \{ \Heart^2 + k(N, d) h^{-1} \Heart + h^{-2} \triangle_{l} \} + V_{l} +  J\} |\chi\rangle 
\label{Em-TDWE} \mbox{ } . 
\eeq
\noindent One is then using one of eq's (\ref{Sicut}--\ref{Jesu}) to express $h$ as an explicit function of $t^{\se\sm}$.   
This does require invertibility 
%
%
in order to set up the $t^{\se\sm}$-dependent perturbation equation 
%
%
This is as opposed to rather than heavy degree of freedom dependent equation,  which I now denote
$
h = h(t^{\se\sm(\sW\sK\sB)}).
$
Even for 1 $h$ variable, this is not in general guaranteed, but the examples in question do possess it.  
The inversion can also be used to convert $h$-derivatives to $t^{\se\sm(\sW\sK\sB)}$-derivatives, so one has a bona fide $l$-equation.

(\ref{Em-TDWE}) is usually approximated by a semiclassical emergent TDSE, 
\beq
i\hbar\Last{\pa\left|\chi\right\rangle} = H_{l}|\chi\rangle =
- \{\hbar^2/2\}{\triangle_{\sfS(N, d)}    }\left|\chi\right\rangle/{    h^2(t^{\se\sm})    }  + V_{l}\left|\chi\right\rangle \mbox{ } . \label{SETDSE}
\eeq
(See Sec \ref{Semi-Avs} for various possible justifications of the approximations made.)   
\noindent (\ref{SETDSE}) is, modulo the $h$--$l$ coupling term, `ordinary relational $l$-physics'.  
In turn, this is `ordinary $l$-physics' modulo the effect of the angular momentum correction term. 
Thus the purported simple situation has `the scene set' by the $h$-subsystem for the $l$-subsystem to have dynamics. 
This dynamics is furthermore slightly perturbed by the $h$-subsystem, while neglecting the back-reaction of the $l$-subsystem on the $h$-subsystem.

\subsection{Use of rectified time, and that this amounts to working on shape space}

Provided that one is focussing on the TDSE core, rather than a more general TDWE form, (\ref{SETDSE}) this further simplifies if one chooses the {\sl emergent rectified time} 
\cite{SemiclI} given by 
$
h^2\Last := h^2 \pa/\pa t^{\se\sm(\sW\sK\sB)} = \pa/\pa t^{\se\sm(\sr\se\scc)} =: \mbox{\textcircled{$\star$}}.  
$
We define this to arbitrary order, though firstly we consider the zeroth order version, i.e. 
\beq
\mbox{\large $t$}^{\se\sm(\sr\se\scc)}_{(0)} =  
\int\d t^{\se\sm(\sW\sK\sB)}_{(0)}/h^2(t^{\se\sm(\sW\sK\sB)}_{(0)}) = 2^{-1/2} \int {\d h}\left/ {h^2\sqrt{{E_h - V(h)}}}\right.   
\hspace{1in} \mbox{ ($h$ = $\rho$) } . 
\eeq
\noindent Once correction terms to which the $l$-physics contributes are included, rectified time is clearly as Machian as emergent WKB time [i.e. also of the form in eq (\ref{hdh})]. 
In fact, the two are related by a conformal transformation, which is a relationally-motivated freedom \cite{Banal}. 
Thus they lie within the same theoretical scheme from the Machian perspective.  
This suggests that, whilst emergent WKB time follows on as a quantum-corrected form of emergent JBB time, 
the mathematics of the quantum system dictates passage to the rectified time instead as regards semiclassical quantum-level calculations.
Using $t^{\sr\se\scc}$ amounts to studying the $l$-physics by working on the shape space itself, i.e. the geometrically natural presentation.     
\noindent $\lt^{\se\sm(\sr\se\scc)}$ is very similar to the geometrically-natural time from the perspective of shape space.  
The difference lies in that the geometrically-natural $\5Star :=  I { \sqrt{W}\d }/{ \d s_{\ts\th\ta\tp\te} }$, whilst \textcircled{$\star$} := $I { \sqrt{W}\d }/{\d s}$,             
Thus both carry the same conformal factor, $I$, but differ as regards the type of the kinetic term involved.                                                           
Finally, if $\lt^{\se\sm(\sW\sK\sB)}$ is monotonic, it is straightforward to show that $\lt^{\se\sm(\sr\se\scc)}$ is too. 

The full rectified $l$-TDWE is then
\beq
i\hbar\{1 - \mP_{\chi}\} \Club |\chi\rangle = 
\{1 - \mP_{\chi}\} \{ - \{\hbar^2/2\} \{\Spade^2 + k(N, d) \Spade  + \triangle_{l} \} +  V^{\sr\se\scc}_{l} + J^{\sr\se\scc}  \}   |\chi\rangle \mbox{ } . 
\label{Rec-TDSE}
\eeq
Here $\Club    := \Rec  - \Rec  l\pa_{l}$  and $\Spade := \Club/\Rec  \, \mbox{ln} \, h(t^{\sr\se\scc})$ are the equation-simplifying groupings of derivatives.  
\noindent This is most commonly viewed as 
perturbations about a TDSE, 
\beq
i\hbar\mbox{\textcircled{$\star$}}|\chi\rangle = -\{\hbar^2/2\}\triangle_{l}\left|\chi\right\rangle + 
V^{\sr\se\scc}_{l} \left|\chi\right\rangle + {J}^{\sr\se\scc}\left|\chi\right\rangle \mbox{ \{+ further perturbation terms\} } . 
\eeq
\noindent Note 1) The rectified time's simplification of the emergent-TDSE equation can be envisaged as passing from the emergent time that is natural to the whole relational space 
to a time that is natural on the shape space of the $l$-degrees of freedom themselves. 
I.e. to working on the shape space of the $l$-physics itself.    

\noindent Note 2) The $l$-subsystem's simplest time is {\sl not} immediately the one provided by the $h$-subsystem.

\subsection{Rectified $h$-equation}

\noindent We rectify the $h$-equation too, to place the system of equations on a common footing in terms of a single time variable. 
\noindent It is
$$
\{\Rec \, \mbox{ln} \, h\}^2 - 2i\hbar\langle \chi |  \Club \Rec  | \chi \rangle - \hbar^2\{ \langle \chi | \Spade^2 | \chi \rangle 
+ k(N, d) \langle \chi | \Spade | \chi \rangle \} 
$$
\beq
- i\hbar k(N, d) \Rec \, \mbox{ln} \, h + \hbar^2\{ k(\xi) - \langle \triangle_l \rangle = 
2\{E^{\sr\se\scc} -  V_h^{\sr\se\scc} - \langle \chi | V_l^{\sr\se\scc} | \chi \rangle - \langle \chi | J^{\sr\se\scc} | \chi \rangle \}\mbox{ } .  
\eeq
Its simplest truncation is 
\beq
\{\Rec \, \mbox{ln} \, h\}^2   = 2\{E^{\sr\se\scc} -  V_h^{\sr\se\scc} \} \mbox{ } \mbox{ } , \mbox{ } \mbox{ which integrates to }  \mbox{ } \mbox{ }  
\lt^{\se\sm(\sr\se\scc)} = \int \d h\left/h\sqrt{2W^{\sr\se\scc}}\right. = \int {\d h}\left/{h^2\sqrt{2W_h}}\right. \mbox{ } .  
\eeq
\noindent Caveat.  The $V_l$--$J$ split is not preserved by the rectifying operation. 
Thus in subsequent working it is not necessarily clear how much of this should carry an $\epsilon$.  
$V^{\sr\se\scc} + J^{\sr\se\scc}$ denotes $h^2\big(\lt^{\sr\se\scc}_{(0)}\big)\{V_{l} + J\}$.   
%
%
\noindent The $J/V_l$ ratio is however preserved for those pieces that do not change identity from $J$ to $V_l$ or vice versa.
\noindent Note that specific RPM examples of rectified emergent TDSE's 
%
%
are mathematically familiar equations \cite{FileR}. 
They are TDSE counterparts of \cite{FileR}'s TISE's.  
%

\subsection{Types of contribution to the Machian semiclassical time}

\noindent Expanding out and keeping up to 1 power of $\hbar$ 
%
\beq
\lt^{\se\sm(\sr\se\scc)} = \lt^{\se\sm(\sr\se\scc)}_{(0)} + \frac{1}{2\sqrt{2}}\int\frac{\langle\chi| J |\chi\rangle}{W_{h}^{3/2}}\frac{\d h}{h^2} - 
\frac{i\hbar}{4}  \int   \frac{\d h}{h^2 W_{h}}  \left\{  \frac{k(N, d)}{h} + 2\langle\chi|\pa_{h}|\chi\rangle \right\} + O(\hbar^2) \mbox{ } .   
\label{QM-expansion}
\eeq
I.e., with comparison with the classical counterpart (\ref{Cl-Expansion}) an `expectation of interaction' term in place of an interaction term, 
and an operator-ordering term and an expectation term in place of a classical $l$-change term.   

Note that the first correction term can be interpreted in terms of an $F_q :=\langle J \rangle/W_h$. 
The classical use of a judging criterion $F_{\sj}$ should in some sense carry over to this semiclassical working.  

\noindent Regime 0) Even if expectation terms are small, there is a novel operator-ordering term. 
Incorporating this does {\sl not} require coupling the chronifer procedure to the quantum $l$-equation.  
It is a quantum correction to the nature of the scale physics itself rather than a Machian $l$-subsystem change contribution.  
\noindent This working suffices to show that the Machian emergent time finding procedure can return {\sl complex} answers in the semiclassical, and, more generally, fully quantum, regimes.  
This opens up questions of interpretation.
(See also e.g. basic QFT \cite{basicQFT}, Complex Methods complications in curved spacetime \cite{HL90}, and Bojowald et al.'s recent work \cite{Bojo}.)
Complex entities are common enough in quantum theory (e.g. slightly deformed contour integrals in expressions for propagators). 
However what complex methods are well-established to work in flat geometries encounter further difficulties in passing to curved-geometry cases required by GR.  
\noindent This correction term is readily evaluable for some simple examples. 
It is $i\hbar k(N, d)/2E_{\sU\sn\si}h^2$ in the free shape momentum $\fS = 0$ case, and, in in the HO $\fS = 0$ case,
\beq
\frac{i\hbar k(N, d)}{4E_{\sU\sn\si}}
\left\{ 
\frac{1}{2h^2} + \frac{A}{E_{\sU\sn\si}}\mbox{ln}
\left(
\frac{\sqrt{E_{\sU\sn\si} - Ah^2}}{h}
\right)
\right\} \mbox{ } . 
\eeq

\noindent To explicitly evaluate the other two terms here, we need coupling to the $l$-equation to have $|\chi\rangle$ (see Sec \ref{Det-Back}).
The above expansion suffices, however, to demonstrate the Machian character of the emergent WKB time now indeed give the QM $l$-subsystem an opportunity to contribute:
\beq
\lt^{\se\sm(\sW\sK\sB)} = {\cal F}[ h, l, \d h, |\chi(h, l) \rangle ] \mbox{ } .  
\eeq
\noindent In greater generality than the above $\hbar$ expansion,
\beq
\mbox{\Large $t$}^{\se\sm(\sr\se\scc)} \propto \int {2\,\d h}\left/{h^2\left\{  -B \pm \sqrt{B^2 - 4C}\right\}}\right.    \mbox{ } ,  
\eeq
\beq
B := - i\hbar\{2 \langle\chi|\pa_{h}|\chi\rangle +  h^{-1}k(N, d)\} \mbox{ } , \mbox{ } \mbox{ } 
C := -2\{W_{h} - \langle\chi| V_l + J |\chi\rangle\} + \hbar^2\{   h^{-1}k(N, d)\langle\chi|\pa_{h}|\chi\rangle - \langle\chi|\pa_{h}^2|\chi\rangle + h^{-2}c(N, d)\} \mbox{ } .  
\eeq
Thus what were pairs of solutions differing only by a $\pm$ sign at the classical level are turned into more distinct complex pairs. 
This splitting is mediated by operator-ordering and expectation contributions to first order in $\hbar$. 
One also sees that the second-order contributions are another expectation, another ordering term and one that has one factor's worth of each.

\subsection{Some simple $l$-TDSE regimes}

\noindent Overall, the fluctuation $l$-equation (\ref{l-TDSE-prime}) can be rearranged to obtain a TDSE with respect to an emergent time `provided by the $h$-subsystem'. 
\noindent The main idea is then to consider (\ref{TimeSet}) and (\ref{l-TDSE-prime}) as a pair of equations for the unknowns $t^{\se\sm(\sr\se\scc)}$ and $|\chi\rangle$.

\noindent Regime 1)  One might argue for the interaction term $\langle J\rangle$ to be quantitatively negligible as regards the observed $l$-physics, as both a small interaction term 
and an averaged quantity.   

\noindent Regime 2)  Instead keeping this interaction term, then (\ref{Rec-TDSE}) has not only a time provided by the $h$-subsystem but also a time-dependent imprint on the 
$l$-subsystem's physics due to the $h$-subsystem's physics.  
I.e. neglect the averaged terms and the unaveraged first and second derivative terms (see Secs \ref{Semi-Avs} and \ref{Semi-Deri} for various possible justifications).   
\noindent Thus Regime 1) amounts to solving a Hamilton--Jacobi equation and then a non-interacting TDSE.  
On the other hand, Regime 2 amounts to solving a Hamilton--Jacobi equation and then an interacting TDSE (e.g. as a time-dependent perturbation about the non-interacting TDSE).  
\noindent Regimes 1b) and 2b) extend these two systems by allowing for back-reaction of the $l$-subsystem on the $h$-subsystem, via e.g. the $h$-equation including the term $h$4): 
$\langle\chi|J|\chi\rangle$.  
Only the case with interaction {\sl and} back-reaction makes detailed and/or long-term sense from the perspective of energy-balance `book-keeping'.  
I.e. that, energy transitions in the one system have to be compensated by opposing energy transitions in the other subsystem.

\subsection{Backreaction terms}\label{Semi-Back}

\noindent One interesting feature is that the $l$-subsystem can back-react on the $h$-subsystem rather than just merely receive a time-standard from it 
(see \cite{BV89}, or \cite{Kieferbook} for a review).
The $l$-equation is now coupled to a less approximate chroniferous $h$-equation containing operator ordering and expectation quantum terms.
Here, the perturbations of expectation type having input from the $l$-subsystem (they are expectation values in the $l$-subsystem's wavefunction).  
Clearly then the previously-suggested simple procedure of solving the $h$-Hamilton--Jacobi equation first is insufficient by itself to capture this level of detail.  

As a first motivation, the current paper's scheme does allow for such terms. 
Moreover, it points out the significance of the expectation term corrections to the $h$-equation as implementing Mach's Time Principle in a STLRC way.  
This gives the $l$-subsystem the opportunity to contribute to the final more accurate estimate of the emergent timefunction.  
\noindent The presence of these corrections makes further physical sense in accord with the following second motivation. 
\noindent The Hamilton--Jacobi equation approximation here depicts a conservative system.
But if the $h$-subsystem interacts with the $l$-subsystem one is to expect it to have a more general form than the conservative one 
Then indeed, expectation terms can be seen as functionals of $|\chi(l^{\sfa}, t^{\se\sm(\sW\sK\sB)}_{(0)})\rangle$ with the integration involved indeed 
not removing the $t^{\se\sm(\sr\se\scc)}_{(0)}$ dependence.  
Thus the $h$-equation containing these corrections is indeed dissipative rather than conservative.
A third motivation is the preceding SSec's book-keeping argument.
A fourth motivation is that {\sl back-reaction is conceptually central to GR}.  
(This both at the level of what the Einstein field equations mean and in GR's aspect as supplanter of absolute structure.) 
Thus model arenas that include back-reaction are conceptually desirable in schemes that concentrate on better understanding GR.  
Note that these regimes just involve time-dependent perturbations of standard simple TDSE's (for all that these two uses of `time' are now interpreted as `emergent time').
See \cite{SemiclI, SemiclIII} for previous simple examples of backreaction.

\subsection{Detail of the small but non-negligible back-reaction}\label{Det-Back}  

In the case that $J \approx 0$ suffices in the $l$-equation, the other two low-order terms in (\ref{QM-expansion}) are 1) 
$$
- \frac{i\hbar}{2}         \int  \frac{\d h}{h^2} \frac{\langle\chi| \pa_{h} |\chi\rangle}{W_{h}} =  
- \frac{1}{2\sqrt{2}} \int  \frac{\d h}{h^2} \, \frac{\hbar^2\md^2/2 + Ah^4}{\{E_{\sU\sn\si} - Ah^2\}^{3/2}} = 
$$
\beq
  \frac{1}{2\sqrt{2}}
\left\{
\frac{1}{\sqrt{E_{\sU\sn\si} - Ah^2}}
\left\{
\frac{\hbar^2\md^2}{2E_{\sU\sn\si}}
\left\{
\frac{1}{h} - \frac{2A}{E_{\sU\sn\si}}h
\right\}
- h
\right\}
+
\frac{1}{\sqrt{A}}
\mbox{arctan}
\left(
\frac{\sqrt{A}h}{\sqrt{E_{\sU\sn\si} - Ah^2}}
\right)
\right\}
\mbox{ } . 
\label{correction-1b}
\eeq
See \cite{QuadIII} for parallel treatment of the $\langle\pa_h^2\rangle$ correction term.  
\beq
\frac{1}{2\sqrt{2}}\int\langle\chi| J |\chi\rangle \frac{\d h}{h^2W_{h}^{3/2}} = \frac{B\langle d| \mbox{cos}^2\varphi |d \rangle}{2\sqrt{2}}  \int \frac{\d h}{h^2W_{h}^{3/2}} 
                                                                               = \frac{Bh}{2\sqrt{2}E_{\sU\sn\si}\sqrt{E_{\sU\sn\si} - Ah^2}}  \mbox{ } .  
\hspace{1in} 
\eeq 
These are for 3-stop metroland, but it is not hard to generalize them to further RPM's \cite{QuadIII}. 
(\ref{correction-1b})'s integral remains in terms of basic functions by virtue of $v = \rho^2$ substitution and completing the square. 

For a more in-depth treatment in terms of expansions -- using $t = t_{(0)} + \epsilon t_{(1)}$ and $|\chi\rangle = |\chi_{(0)}\rangle + \epsilon|\chi_{(1)}\rangle$, see \cite{SemiclIII}.  
Here the $\epsilon$ originally comes from being a split-out factor in front of the interaction term $J$.  
Usually the first $J$ should be kept, since elsewise the $l$-subsystem's energy changes without the $h$-system responding, violating conservation of energy.    
But if this is just looked at for a ``short time" (few transitions), the drift may not be great. 
It may lie within the uncertainty to which an internal observer would be expected to know their universe's energy.  
Here, I do not explicitly perturbatively expand the last equation as it is a decoupled problem of a standard form.
I.e. a $\lt^{\se\sm(\sr\se\scc)}_{(0)}$-dependent perturbation of a simple and well-known $\lt^{\se\sm(\sr\se\scc)}_{(0)}$-dependent perturbation equation.
This scheme can be solved in terms of Green's Functions \cite{SemiclIII}.

\subsection{Averaged terms}\label{Semi-Avs}

Expectation/averaged terms are often dropped in the Quantum Cosmology literature.
The usual line given for this in that literature is that these are argued to be negligible by the 
\noindent{\it Riemann--Lebesgue Theorem}, which is the mathematics corresponding to the physical idea of {\it destructive interference}.  


I add that Quantum Cosmology practitioners probably do not want such terms to be around due to non-amenability to exact treatment that they confer upon the equations if included.  
However, here are some reasons to keep it.   


\noindent 1) Some RPM counterexamples to these terms being small are as follows. 
For 3-stop metroland's analogue of the central problem, $\langle\pa_{\varphi}^2\rangle|\chi\rangle$ and $\pa_{\varphi}^2|\chi\rangle$ are of the same size since 
the wavefunctions in question are eigenfunctions of this operator.
%

\noindent 2) Moreover, then $H_l|\chi\rangle = \{\triangle_l - \langle\triangle_l\rangle\}|\chi\rangle$ gives zero rather than $\triangle_l|\chi\rangle$.
Still, the solution to the unaveraged equation solves the averaged equation too, and constitutes a proper eigenfunction (unlike 0).  
This approach suggests keeping all average terms in the $l$-equation together.

\noindent 3) I have pointed out \cite{SemiclIII} an analogy with Atomic/Molecular Physics, where the counterparts of such terms require a 

\noindent {\it self-consistent} variational--numerical approach.  

\noindent An example of this is the iterative technique of the {\it Hartree--Fock approach}. 
In Atomic/Molecular Physics, it is is then conceded that this ensuing non-exactly tractable mathematics is necessary so as to get passably correct answers (experimentally confirmed).  
I investigate the quantum-cosmological counterpart of this in more detail in \cite{SemiclIV}.  

\noindent While there are a number of differences between Molecular Physics and Quantum Cosmology, Hartree--Fock theory in fact is known to span those differences.
E.g. it is available for time-dependent physics, and involving a plain rather than antisymmetrized wavefunction, and for field theory (c.f. Condensed Matter Physics \cite{Condi}).  


\noindent Regime 3) $t^{\se\sm(\sr\se\scc)}_{(0)}$ is satisfactory, then apply a Hartree--Fock type procedure on the $l$-equation with average terms kept.  
%
%
\noindent Such a procedure requires variational justification, which is covered in \cite{SemiclIV}.
\noindent However, this is insufficient if one's scheme is to comply with Mach's Time Principle.
\noindent It is then not as yet clear how to extend the self-consistent treatment to this non-negligible back-reaction case (Regime 3b).
Schematically, this is of the form 
\beq
\mbox{\Huge\{}  \stackrel{\mbox{( chroniferous Hamilton--Jacobi equation with expectation corrections)}}
                         {\mbox{(emergent-time-dependent Hartree--Fock scheme)}}
                          \mbox{ } , 
\label{cyril}  
\eeq
which is probably on this occasion a new type of system from a Mathematical Physics perspective.  
Can this be anchored to a variational principle? 
Thus this investigation does not just concern qualitative confidence in the Halliwell--Hawking scheme 
but is also important as regards the detailed robustness of the Semiclassical Approach's time-emergence itself.

\subsection{Higher derivative terms}\label{Semi-Deri}

One often neglects the extra $t^{\se\sm(\sr\se\scc)}$-derivative terms whether by discarding them prior to noticing they can be converted into $t^{\se\sm(\sr\se\scc)}_{(0)}$-derivatives 
or by arguing that $\hbar^2$ is small or $\rho$ variation is slow. 
Moreover there is a potential danger in ignoring higher derivative terms even if they are small (c.f. Navier--Stokes equation versus Euler equation in fluid dynamics).
\noindent One would expect some regions of configuration space where the emergent TDWE behaves more like a Klein--Gordon equation than a TDSE, albeit in full it is more general.    
Thus the guarantee of appropriate interpretability for TDSE's is replaced by a more difficult study of a more general TDWE.  
%
%
\noindent Kiefer and Singh's expansion \cite{KS91} treats higher derivative terms along the lines of the next-order correction to the TDSE from the Klein--Gordon equation.

\subsection{Semiclassical emergent Machian time: perturbative scheme}

\noindent The logical conclusion of using equation-simplifying time leads one to formulating the Semiclassical Approach for scaled RPM's in terms of $t^{\se\sm(\sr\se\scc)}$.  
This is a manifestation of  Regime 3): 
solve for $t = t(h, l^{\sfc})$: no independent notion of time and for $|\chi(l, t^{\se\sm(\sr\se\scc)}\rangle$ (standard QM for the $l$-subsystem with respect to the emergent time).
As these are the functions to solve for, they are to be perturbed.  
Classically the $Q$'s are perturbed and this is required here since $t^{\se\sm(\sr\se\scc)}$ is in terms of them.
This differs then from standard QM perturbation theory in which the $Q$'s are not perturbed.  
All in all, we now take 
\beq
Q^{\sfA} = Q^{\sfA}_{(0)} + \underline{\epsilon}\cdot\underline{Q}^{\sfA}_{(1)} + O(\epsilon^2) \mbox{ } , \mbox{ for both $h$ and $l$, }
\label{B-1}
\eeq
\beq
t^{\se\sm(\sr\se\scc)} = t^{\se\sm(\sr\se\scc)}_{(0)} + \underline{\epsilon}\cdot\underline{t}^{\se\sm(\sr\se\scc)}_{(1)} + O(\epsilon^2) \mbox{   and    }
\label{B-2}
\mbox{ } 
|\chi\rangle = |\chi_{(0)}\rangle +  \underline{\epsilon}\cdot |\underline{\chi}_{(1)}\rangle + O(\epsilon^2) \mbox{ } .
\eeq
This amounts to simultaneous consideration of Sec \ref{+temJBB}'s perturbations and the current Sec's perturbations in a Mach's Time Principle context.
(Both are given the opportunity to contribute to $t^{\se\sm(\sr\se\scc)}$ perturbations.)  

Note that for some purposes (a set of relevant $\epsilon_u$) some of the corresponding responses (e.g. $t_{(1)u}$, $l_{(1)u}$, $q_{(1)u}$) would be expected to be negligible.
For instance, we can turn on a small electric field in our laboratory to study the Stark Effect in atoms without significantly affecting the timestandard.  
Once we are sure this is the case for a particular set-up, it can be modelled by a rather less all-encompassing set of perturbed quantities than is considered above.
The full system would only be used for quantum-cosmological applications in which an accurate emergent time is required.

Then use each term's label to also label the corresponding $\epsilon$ to obtain  
$$ 
\frac{\{\Rec h\}^2}{h^2} - 
2i\epsilon_a  \left\{\frac{\hbar}{\epsilon_a} \langle\chi|   \frac{\{    \Rec\!-\!\epsilon_i\{\Rec l \pa_{l}/\epsilon_i\}}{h^2}  \} |\chi\rangle \right\} - 
\left.
\epsilon_b \mbox{\LARGE $\langle$} \chi \mbox{\LARGE $|$}\frac{\hbar^2}{\epsilon_b} 
\mbox{\LARGE $\{$}\frac{\Rec  - \epsilon_i\{\Rec l\pa_{l}/\epsilon_i\}}{\Rec h}\mbox{\LARGE $\}$}^2   \mbox{\LARGE $|$}\chi\mbox{\LARGE $\rangle$}\!-\! +
\frac{k(N, d)}{h} 
\right\{
\epsilon_c  
\mbox{\LARGE $\{$}\frac{\hbar^2}{\epsilon_c} 
\mbox{\LARGE $\langle$}\chi\mbox{\LARGE $|$}
\frac{\Rec\!-\!\epsilon_i\{\Rec l \pa_{l}/\epsilon_i\}}{\Rec h}\mbox{\LARGE $|$}\chi\mbox{\LARGE $\rangle$}  \mbox{\LARGE $\}$}
$$
\beq
\left.
+ \epsilon_d \mbox{\LARGE $\{$} \frac{i\hbar}{\epsilon_d} {\Rec h} \mbox{\LARGE $\}$}
\right\}
+ \epsilon_e
\left\{
\frac{\hbar^2}{\epsilon_e} c(N, d)
\right\} 
- \epsilon_f  
\left\{ 
\frac{\hbar^2}{\epsilon_f}  \langle\chi|\triangle_{l}|\chi\rangle  
\right\}
+ 2V^{\sr\se\scc}_{h}(h) 
+ 2\epsilon_g  \langle\chi|  V^{\sr\se\scc\,\prime}              |\chi\rangle 
+ 2\epsilon_h  \langle\chi|  J^{\sr\se\scc\,\prime}  |\chi \rangle 
= 2E^{\sr\se\scc} 
\mbox{ } \mbox{ } ,   
\label{epsi-TimeSet}
\eeq
$$
2i\hbar \{\Rec - \epsilon_i\{\Rec l \pa_{l}/\epsilon_i\}\}|\chi\rangle 
- \epsilon_q 
\left\{
\frac{2i\hbar}{\epsilon_q} \mP_{\chi} \{\Rec - \epsilon_i\{\Rec l \pa_{l}/\epsilon_i\}\}|\chi\rangle 
\right\} = 
- \epsilon_r h^2
\left\{
\frac{\hbar^2}{\epsilon_r}
\left\{
\frac{\Rec - \epsilon_i\{\Rec l \pa_{l}/\epsilon_i\}}{\Rec h}
\right\}^2     |\chi\rangle 
\right\}
$$
$$
- \epsilon_s h^2
\left\{
\frac{\hbar^2}{\epsilon_s}   \frac{k(N, d)}{h}  \frac{\Rec - \epsilon_i\{\Rec l \pa_{l}/\epsilon_i\}}{\Rec h}  |\chi\rangle 
\right\} 
- \hbar^2  \triangle_{l} |\chi\rangle + 2V^{\sr\se\scc}_{l}|\chi\rangle + 2\epsilon_u J^{\sr\se\scc\,\prime}|\chi\rangle + \epsilon_v h^2
\left\{
\frac{\hbar^2}{\epsilon_v} \mP_{\chi} 
\left\{
\frac{\Rec - \epsilon_i\{\Rec l \pa_{l}/\epsilon_i\}}{\Rec h}
\right\}^2     
|\chi\rangle 
\right\}
$$
\beq
\left. 
+\epsilon_w h^2
\left\{
\frac{\hbar^2}{\epsilon_w}  \frac{k(N, d)}{h}  \mP_{\chi}  \frac{\Rec - \epsilon_i\{\Rec l \pa_{l}/\epsilon_i\}}{\Rec h}  |\chi\rangle 
\right\}
- \epsilon_x 
\left\{
\frac{\hbar^2}{\epsilon_x}  \mP_{\chi} \triangle_{l}  |\chi\rangle
\right\} 
- 2\epsilon_y  \mP_{\chi}  V^{\sr\se\scc\,\prime}_{l}     |\chi\rangle 
- 2\epsilon_z  \mP_{\chi}  J^{\sr\se\scc\,\prime}    |\chi\rangle 
\right\}
\mbox{ } . 
\label{epsi-Em-TDSE}
\eeq
\noindent \mbox{ } \mbox{ } Then zeroth order simply returns
\beq
\{\Rec_{(0)}h_{(0)}\}^2 = 2 h^2 \{E^{\sr\se\scc} - V^{\sr\se\scc}_{h}(h_{(0)})\} \mbox{ } ,
\eeq
\beq
i\hbar\Rec_{(0)}|\chi_{(0)}\rangle = - \{\hbar^2/2\}\triangle_{l_{(0)}} |\chi_{(0)}\rangle + V^{\sr\se\scc}_{l}\big(l^{\sfc}_{(0)}\big) |\chi_{(0)}\rangle \mbox{ } .  
\eeq
\noindent On the other hand, the first-order equations are now 
\noindent
$$
\frac{\underline{\epsilon}}{h_{(0)}^2} \cdot 
\left\{
\Rec_{(0)}h_{(0)}
\left\{
\Rec_{(0)}\underline{h}_{(1)} - \Rec_{(0)}\underline{t}_{(1)}^{\sr\se\scc} 
\right\} 
+ \underline{h}_{(1)}
\left\{
\Rec_{(0)}h_{(0)} + \frac{\pa V^{\sr\se\scc}_{h}}{\pa h_{(0)}} - \frac{2E^{\sr\se\scc}}{h_{(0)}} 
\right\}
\right\} = 
$$
$$
i\epsilon_a
\left\{
\frac{\hbar}{\epsilon_a h_{(0)}^2} \langle \chi_{(0)} |  \Rec_{(0)} | \chi_{(0)} \rangle 
\right\} 
+ \epsilon_b
\left\{
\frac{\hbar^2}{2\epsilon_b}
\langle \chi_{(0)} |   
\left\{ 
\frac{\Rec_{(0)}}{\Rec_{(0)}h_{(0)}}
\right\}^2
| \chi_{(0)} \rangle 
\right\}
+ \frac{K(N, d)}{2h_{(0)}}
\left\{
\epsilon_c
\left\{
\frac{\hbar^2}{\epsilon_c}
\langle \chi_{(0)} |   \frac{\Rec_{(0)}}{\Rec_{(0)}h_{(0)}} | \chi_{(0)} \rangle 
\right\}
\right.
$$
\beq
\left.
\!+\!\epsilon_d
\left\{
\frac{i\hbar}{2\epsilon_d} \frac{\Rec_{(0)}h_{(0)}}{h_{(0)}^2}  
\right\}
\right\}
- \epsilon_e
\left\{ 
\frac{\hbar^2 k(\xi)}{\epsilon_e}
\right\}
\!+\!
\epsilon_f
\left\{ 
\frac{\hbar^2}{2\epsilon_f}
\langle \chi_{(0)} |   \triangle_{l_{(0)}} | \chi_{(0)} \rangle 
\right\}
\!-\!\epsilon_g  \langle \chi_{(0)} |  V^{\sr\se\scc\,\prime}_l(l_{(0)}^{\sfq}) | \chi_{(0)} \rangle 
\!-\!\epsilon_h  \langle \chi_{(0)} |  J^{\sr\se\scc\,\prime}(h_{(0)}, l_{(0)}^{\sfq}) | \chi_{(0)} \rangle 
               \mbox{ } ,  
\eeq
$$
\underline{\epsilon}\cdot
\mbox{\Huge\{}
i\hbar
\left\{
\Rec_{(0)}|\underline{\chi}_{(1)}\rangle - \Rec_{(0)}\underline{t}^{\sr\se\scc}_{(1)}| \chi_{(0)} \rangle 
\right\}
+ \frac{\hbar^2}{2}
\left\{
\triangle_{l_{(0)}}|\underline{\chi}_{(1)}\rangle 
\right\} 
- V_{l}^{\sr\se\scc}\big(l^{\sfq}_{(0)}\big) |\underline{\chi}_{(1)}\rangle  
- \underline{l}_{(1)}^{\sfc} \frac{\pa V \big(l^{\sfq}_{(0)}\big)}{\pa l^{\sfc}_{(0)}}  |\chi_{(0)}\rangle
$$
$$
- \frac{\hbar^2\underline{l}_{(1)}^{\sfp}}{2M(l_{(0)}^{\sfq})}  \frac{\pa M(l_{(0)}^{\sfq})}{\pa l_{(0)}^{\sfp}} \triangle_{l_{(0)}}|\chi_{(0)}\rangle 
- \hbar^2\frac{\pa\underline{l}_{(1)}^{\sfe}}{\pa l^{\sfc}_{(0)}}  \frac{1}{\sqrt{M\big(l_{(0)}^{\sfq}\big)}}  \frac{\pa}{\pa l^{\sfe}_{(0)}}
\left\{ 
\sqrt{M\big(l_{(0)}^{\sfq}\big)}  N^{\sfc\sfd}\big(l_{(0)}^{\sfq}\big)  \frac{\pa }{\pa l^{\sfd}_{(0)}}
\right\}
$$
$$
+ \frac{\hbar^2}{\sqrt{M\big(l_{(0)}^{\sfq}\big)}}\frac{\pa}{\pa l_{(0)}^{\sfc}}
\left\{
\sqrt{M\big(l_{(0)}^{\sfq}\big)}
\left\{
N^{\sfc\sfd}\big(l_{(0)}^{\sfq}\big)
\left\{
\frac{\underline{l}_{(1)}^{\sfs}}{2M\big(l_{(0)}^{\sfq}\big)} \frac{\pa M\big(l_{(0)}^{\sfq}\big)}{\pa l_{(0)}^{\sfs}} \frac{\pa }{\pa l^{\sfd}_{(0)}} - 
\frac{\pa \underline{l}_{(1)}^{\sff}}{\pa l_{(0)}^{\sfd}} \frac{\pa}{\pa l_{(0)}^{\sff}}
\right\}
+ \frac{\pa  N^{\sfc\sfd}\big(l_{(0)}^{\sfq}\big)}{\pa l_{(0)}^{\sfs}}  \underline{l}_{(1)}^{\sfs} \frac{\pa}{\pa l^{\sfd}_{(0)}   }
\right\}
\right\} |\chi_{(0)}\rangle  \mbox{\Huge\}} = 
$$
$$
i\hbar
\left\{
\epsilon_i
\left\{
\frac{\Rec_{(0)}  l_{(0)}^{\sfc} \pa_{l_{(0)}^{\sfc}} |\chi_{(0)}\rangle}{\epsilon_i}
\right\}
+ \epsilon_q
\left\{
\frac{\mP_{\chi_{(0)}}\Rec_{(0)} |\chi_{(0)}\rangle}{\epsilon_q}
\right\}
\right\}
- \epsilon_r h_{(0)}^2
\left\{
\frac{    \hbar^2    }{    2\epsilon_{r}    }  
\left\{     
\frac{   \Rec_{(0)}   }{    \Rec_{(0)}h_{(0)}    }  
\right\}^2 
|\chi_{(0)}\rangle
\right\}
$$
$$
- \epsilon_s h_{(0)}
\left\{
\frac{\hbar^2}{2\epsilon_{s}} {k(N, d)} \frac{\Rec_{(0)}}{\Rec_{(0)}h_{(0)}} |\chi_{(0)}\rangle
\right\}
+ \epsilon_u J^{\sr\se\scc\,\prime}(h_{(0)}, l_{(0)}^{\sfq}) |\chi_{(0)}\rangle + \epsilon_v h_{(0)}^2 \mP_{\chi_{(0)}}  
\left\{
\frac{\hbar^2}{2\epsilon_{v}}\left\{  \frac{\Rec_{(0)}}{\Rec_{(0)}h_{(0)}}  \right\}^2 |\chi_{(0)}\rangle
\right\}
$$
\beq
+\!\epsilon_w h_{(0)} \mP_{\chi_{(0)}}
\left\{
\frac{\hbar^2}{2\epsilon_{w}}   k(N, d)   \frac{\Rec_{(0)}}{\Rec_{(0)}h_{(0)}} |\chi_{(0)}\rangle
\right\}\!-\!\epsilon_{x} \mP_{\chi_{(0)}}
\left\{
\frac{\hbar^2}{2\epsilon_x} \triangle_{l_{(0)}}|\chi_{(0)}\rangle 
\right\}\!-\!
\epsilon_y \mP_{\chi_{(0)}} V_l^{\sr\se\scc\,\prime}(l_{(0)}^{\sfq}) |\chi_{(0)}\rangle\!-\!\epsilon_z \mP_{\chi_{(0)}} J^{\sr\se\scc\,\prime}(h_{(0)}, l^{\sfq}_{(0)})|\chi_{(0)}\rangle 
\mbox{ } .
\eeq
\noindent It is then straightforward to specialize this system of equations to the 3-stop metroland and triangleland cases.

\subsection{Other classical Problem of Time facets for classical RPM's}

\noindent The quantum Problem of Beables is approached by selecting a suitable subalgebra of the classical \K beables.  
This is done for $N$-stop metroland and triangleland in \cite{FileR}.
The quantum Constraint Closure Problem is overcome in the r-formulation to these models by straightforward computation \cite{FileR}.
\noindent Sec \ref{Others}'s arguments -- for the rest of the Problem of Time Facets to be absent or not part of the current `a local resolution' claim of the present paper 
-- transcend to the quantum level.

\section{Conclusion}\label{22-end}

Classical and semiclassical schemes are presented that are timeless at the primary level and recover time from Mach's `time is to be abstracted form change' 
principle at the emergent secondary level.
This paper considers this for Relational Particle Mechanics (RPM) model arenas.  
See \cite{AMSS1} for the minisuperspace counterparts; 
Perturbative midisuperspace counterparts of this are forthcoming. 
The classical scheme is Barbour's, cast here explicitly as the classical precursor of the Semiclassical Approach by use of the $h$--$l$ split in the quantum cosmological analogue case. 
(The square root of moment of inertia is $h$ and pure shape is $l$.)

The semiclassical scheme is a Machian variant of the Semiclassical Approach to the Problem of Time (Problem of Time) in Quantum Gravity. 
$t^{\se\sm(\sW\sK\sB)} = t^{\se\sm(\sJ\sB\sB)}$ to zeroth (non-Machian) order.  
They differ to first order.  
Are necessarily distinct, since {\sl quantum} change is part of from where the latter's timestandard is abstracted.  
Moreover, $t^{\se\sm(\sW\sK\sB)}$ is rectified as a second application of equation simplifying.  
See \cite{AMSS1} for a minisuperspace counterpart of the present paper.   
The present paper gives a complete Machian resolution of the classical and semiclassical Problem of Time for 1- and 2-$d$ RPM's, modulo two caveats.

\noindent a) This analysis has not covered the possible need to construct Dirac beables.

\noindent b) This analysis has not justified the crucial WKB ansatz in the first place \cite{Battelle}.  
It is not natural compared to a superposition of such wavefunctions \cite{Zeh88, BS, B93, I93, Kuchar92, EOT, Giu, Zehbook}.  
Justification of WKB in ordinary QM follows from the pre-existence of a surrounding classical large system \cite{LLQM}. 
But this no longer applies for the whole universe. 
Nor does Quantum Cosmology possess ``pure incoming wave laboratory set-up".
Moreover, not being able to justify the WKB ansatz in the Semiclassical Approach to the Problem of Time is a particular problem \cite{Zeh86, Zeh88, BS, Kuchar92, B93, I93, B94II, EOT}.
This is since its its trick by which the chroniferous cross-term becomes the time-derivative part of a TDSE is exclusive to WKB ansatz wavefunctions.

a) and b) are then resolved by a combined (semi)classical--histories--timeless records scheme, as per \cite{AHall, FileR, CapeTown12} 
(built upon the non-Machianly interpreted \cite{H03}). 
This begins with b) being addressed by {\sl decoherence}. 
(Some support for -- but also reservations about -- this have been expressed in e.g. \cite{Kuchar92, I93, Giu, Kiefer93, HT, H03}.) 
N.B. this is quantum-cosmological decoherence, which exhibits some differences from the QM concept \cite{Kiefer99, Kieferbook, Giu}.  
Histories Theory is the most likely source of such decoherence in Quantum Cosmology.  
The question of what decoheres what then leads to consideration of timeless records as well. 
\noindent Then e.g. Halliwell's way of phrasing timeless propositions leads to quantities commuting with Quad. 
If built out of Kucha\v{r} beables, also commute with Lin$_{\sfZ}$ and hence constitute Dirac beables.

I have also given an improved presentation of the semiclassical approach, with qualitative physical analysis of neglected terms and associated regimes of study.  
Some papers \cite{KS91, SCB3, Kiefersugy} investigate Quantum Cosmology by expanding in 1 parameter.  
There are however multiple parameters, as pointed out by Padmanabhan \cite{Pad} and investigated explicitly in the present Article.

\noindent Moreover, the current paper's examples are comparable against outcome of exact quantization (a useful and relatively unusual feature).  
\noindent We included proposing a scheme for a quantum-cosmological generalized local ephemeris time.  
\noindent One might think of this in terms of bare and dressed quantities, though the type of dress is somewhat unusual. 
E.g. it is classical, though fluid mechanics has an effective mass concept too.  
Most of all it is a Machian dress.
\noindent I provided a nontrivial Machian perturbation theory to first order for classical and semiclassical schemes. 

\mbox{ } 

\noindent{\bf Acknowledgements}.  I thank those close to me for support.  
Jeremy Butterfield, Sean Gryb, Philip Hoehn, Pui Ip, Chris Isham, Sophie Kneller, Marc Lachieze-Rey, Matteo Lostaglio, Flavio Mercati, Brian Pitts 
and the Anonymous Referees for discussions.
I was funded by a grant from the Foundational Questions Institute (FQXi) Fund, 
a donor-advised fund of the Silicon Valley Community Foundation on the basis of proposal FQXi-RFP3-1101 to the FQXi.  
I thank also Theiss Research and the CNRS for administering this grant, held at APC Universit\'{e} Paris 7.


\appendix\section{Dirac quantization version of semiclassical schemes}\label{Dir-Semi}

The general case here is, for a shape-nonshape $h$--$l$ split 
\beq
i\hbar
\left\{ 
\frac{\partional}{\partional \lt^{\se\sm(\sr\se\scc)}} - \frac{\partional g^{\sfZ} }{\partional \lt^{\se\sm(\sr\se\scc)}} \widehat{\mbox{Lin}}_{\sfZ}           
             \right\}|\chi\rangle \propto - \frac{\hbar^2}{2}\triangle^{\scc}_{\mbox{\scriptsize Preshape}}|\chi\rangle + ... \mbox{ } . \label{Roth}
\eeq
Her $\partional$ is a partial derivative $\pa$ for finite theories and a functional derivative $\delta$ for field theories.
This is accompanied by 
\beq
\mbox{\lt}^{\se\sm(\sr\se\scc)} \propto \stackrel{\mbox{\scriptsize extremum $\d g$ $\in$ \sfG}}
                                                 {\mbox{\scriptsize of $S_{\ts\te\tm\ti}$}}
\left(
\int ||\d_{g}h||\left/\left\{-B \pm \sqrt{B^2 - 4C}\right\}h^2\right.   
\right)  \mbox{ } .  
\label{ylch}
\eeq
Here the extremization is unnecessary in shape--scale split RPM's and minisuperspace but involving an object $S_{\sss\se\sm\si}$ whose detailed form remains to be specified in 
the next update of \cite{FileR}.  
This unknown object reduces to the relational action $S_{\sr\se\sll}$ in the classical limit but presumably contains quantum corrections.  
$B$ and $C$ are generalizations of the previous specific example of forms for these.
Finally, these equations are further accompanied by $h$- and $l$-Lin$_{\sfZ}$ equations,
\beq
\langle \widehat{\mbox{Lin}}_{\sfZ} \rangle = 0 \mbox{ } , \mbox{ } \mbox{ } \{1 - \mP_{}\chi\}\widehat{\mbox{Lin}}_{\sfZ}|\chi\rangle = 0 \mbox{ } .  
\eeq
The form of (\ref{Roth}) specifically for scaled RPM is then
\beq
i\hbar
\left\{
\frac{\pa}{\pa\lt^{\se\sm(\sr\se\scc)}} - \frac{\pa\underline{B}}{\pa\lt^{\se\sm(\sr\se\scc)}}\cdot\hat{\underline{\cal L}}
\right\}|\chi\rangle = - \frac{\hbar^2}{2}\triangle^{\scc}_{\sfP(N, d)}|\chi\rangle + ... = - \frac{\hbar^2}{2}\triangle^{\scc}_{\mathbb{S}^{nd - 1}}|\chi\rangle + ... \mbox{ } , 
\eeq 
for preshape space $\fP(N, d) := \fQ(N, d)/$Dil, which has the simple geometrical form  $\mathbb{S}^{nd - 1}$ \cite{Kendall}.    
On the other hand, in the GR case the Tomonaga--Schwinger equation,
\beq
i\hbar\left\{\frac{\delta}{\delta\mt^{\se\sm(\sr\se\scc)}} -  {\cal M}_{\mu}\frac{\delta{\mF}^{\mu}}{\delta\mt^{\se\sm(\sr\se\scc)}} \right\}|\chi\rangle = 
\widehat{\mH}_{l}^{\sG\sR}|\chi\rangle \mbox{ }   ,
\label{Tom-Schwi}  \mbox{ } 
\eeq
can be furtherly expressed as the new equation 
\beq
i\hbar
\left\{
\frac{\delta}{\delta{\cal \lt^{\se\sm(\sr\se\scc)}}} - \frac{\delta\underline{\mF}^{\mu}}{\delta \lt^{\se\sm(\sr\se\scc)}}\hat{{\cal M}_{\mu}}
\right\}|\chi\rangle = - {\hbar^2}\triangle^{\scc}_{\sC\sR\si\se\sm(\sbSigma)}                                      |\chi\rangle         + ...   =   
                       - {\hbar^2}\triangle^{\scc}_{\sbU}                          |\chi\rangle                                          + ...       \mbox{ } . 
\eeq
%
Here CRiem($\Sigma$) is the configuration space conformal Riem:  = Riem($\Sigma$)/Conf($\Sigma$), where Conf($\Sigma$) are the conformal transformations. 
Furthermore, $U^{\mu\nu\rho\sigma} := h^{\mu\rho}h^{\nu\sigma}$ is the features DeWitt's \cite{DeWitt67} positive-definite configuration space metric.       
Also, in the GR case, $\lt^{\sr\se\scc}$ bears a slightly different relation to $\lt^{\se\sm(\sW\sK\sB)}$ \cite{SemiclII}. 
This difference is due to the nonuniqueness in radial variables encapsulated by taking, in place of $r$, some $f(r)$ that is monotonic on a suitable range.

\section{Bohmian counterpart}

(\ref{ylch}) requires an `extremum $S_{\sss\se\sm\si}$ at the semiclassical level for the Dirac version of the work which full GR probably requires due to the Thin Sandwich Problem 
aspect of the Problem of Time.   
While a semiclassical suggestion for this will appear in version 4 of \cite{FileR}, the Bohmian approach also has a natural place for semiclassical (and indeed fully quantum-mechanical) 
counterparts of the action.   
This comes hand in hand with the Bohmian interpretation parallelling the classical interpretation in terms of trajectories.

A Bohmian semiclassical approach to Quantum Cosmology is in excess of the ordinary treatment of semiclassicality of Bohmian mechanics on pp 186-191 of \cite{Bohm}.  
Such is considered instead e.g. in \cite{CaWein}
%
%
and also involves a WKB ansatz. 
However, its interpretation is different, since its notion of time has some absolute characteristics. 
%
%
Thus, given semiclassical Quantum Cosmology, there is an interpretational fork that one can take. 
a) Interpreting it in Machian terms in terms of a slight deviation from cosmic time. 
b) Interpreting it in Bohmian partly-absolute format in terms of precisely and privilegedly cosmic time.


\end{document}